\def\aap{A\&A}

\def\apj{ApJ}
\def\apjl{ApJL}

\def\nat{Nature}
\def\mnras{MNRAS}

\def\deg{$^\circ$}

\documentclass[]{aa}  
\usepackage{graphicx}

\usepackage{txfonts}
\usepackage[authoryear,longnamesfirst]{natbib}
\begin{document}
   \title{VLTI monitoring of the dust formation event of the Nova V1280\,Sco\thanks{Based on observations made with the Very Large Telescope Interferometer at Paranal Observatory under programs 278.D-5053, 279.D-5014 and 079.D-0415}
   }

   \author{O. Chesneau
          \inst{1}
          \and
          D.~P.~K. Banerjee\inst{2}
          \and
          F. Millour\inst{3}
          \and
         N. Nardetto\inst{3}
          \and
          S. Sacuto\inst{1}
          \and
          A. Spang\inst{1}
       \and
          M. Wittkowski\inst{4}          
          \and
          N.~M. Ashok\inst{2}
          \and
          R.~K. Das\inst{2}
          \and
          C. Hummel\inst{4}
          \and
          S. Kraus\inst{2}
          \and
          E. Lagadec\inst{5}
          \and
          S. Morel\inst{6}
          \and
          M. Petr-Gotzens\inst{4}
          \and
          F. Rantakyro\inst{6}
          \and
          M. Sch\"oller\inst{6}
          }

   \offprints{O. Chesneau}

   \institute{UMR 6525 H. Fizeau, Univ. Nice Sophia Antipolis, CNRS, Observatoire de la C\^{o}te d'Azur, 
Av. Copernic, F-06130 Grasse, France
              \email{Olivier.Chesneau@obs-azur.fr}
              \and
              Physical Research Laboratory, Navrangpura, Ahmedabad, Gujarat, India
              \and
              Max-Planck-Institut f\"ur Radioastronomie, Auf dem H\"ugel
69, 53121 Bonn, Germany
\and
European Southern Observatory, Karl-Schwarzschild-Strasse 2 D-85748 Garching bei M\"unchen, Germany
\and
Jodrell Bank Center for Astrophysics, The University of Manchester Oxford Street
Manchester M13 9PL, United Kingdom
\and
European Southern Observatory, Casilla 19001, Santiago 19, Chile}

   \date{Received ;accepted }

 
  \abstract
   {We present the first high spatial resolution monitoring of the dust forming nova V1280 Sco performed 
with the Very Large Telescope Interferometer (VLTI).}
   {These observations aim at improving the distance determination of such events and constraining the 
mechanisms leading to very efficient dust formation under the harsh physical conditions encountered 
in novae ejecta.}
   {Spectra and visibilities were regularly obtained from the onset of the dust formation 23 days 
after discovery (or 11 days after maximum) till day 145, using the beam-combiner instruments AMBER (near-IR) and MIDI (mid-IR). These interferometric observations are complemented by near-infrared data from the 1.2m Mt. Abu Infrared Observatory, India. The observations are first interpreted with simple uniform and Gaussian 
models but more complex models, involving a second shell, are necessary to explain the data obtained from 
t=110d after outburst. This behavior is in accordance with the light curve of V1280\,Sco which exhibits  
a secondary peak around t=106d, followed by a new steep decline, suggesting a new dust forming event. 
Spherical dust shell models generated with the DUSTY code are also used to investigate the parameters of 
the main dust shell.  }
   {Using uniform disk and Gaussian models, these observations allow us to determine an apparent linear 
expansion rate for the dust shell of 0.35 $\pm$ 0.03 mas day$^{-1}$ and the approximate time of ejection 
of the matter in which dust formed as t$_{ejec}=10.5\pm7$d, i.e. close to the maximum brightness. This 
information, combined with the expansion velocity of 500$\pm$100km.s$^{-1}$, implies a distance estimate 
of 1.6$\pm$0.4kpc. The sparse $uv$ coverage does not allow to get clear indications of deviation from 
spherical symmetry. The dust envelope parameters were determined. The dust mass generated was typically 
2-8 10$^{-9} M_\odot$ day$^{-1}$, with a probable peak in production at about 20 days after the detection of dust and another peak shortly after t=110d, when the amount of dust in the shell was estimated as 2.2 10$^{-7} M_\odot$. Considering that the dust forming event lasted at least 200-250d, the mass of the ejected material is likely to have exceeded 10$^{-4} M_\odot$. The conditions for the formation of multiple shells of dust are also discussed.   }
   {}

   \maketitle
%

\section{Introduction}
Infrared observations of classical novae have established that dust grains can condense in the 
harsh environment formed in the ejecta generated by thermonuclear 
runaways on the surfaces of white dwarfs in close binary systems \citep{1988ARAA..26..377G,1998PASP..110....3G}.
The formation and evolution of dust grains, following a nova outburst, remains a difficult problem both from 
the observational and the theoretical point of view. The typical light curve during dust formation shows a deep minimum lasting for months; the observed timescale for dust formation is short - typically of the order of 1-20 days - indicating 
an  efficient process of dust formation. Most dust-forming novae condense amorphous carbon which produces
a featureless gray body or blackbody spectral energy distribution (SED). These appear to result from 
thermonuclear runaways on low-mass, carbon-oxygen (CO) WDs that tend to eject more mass under less
energetic conditions than novae on oxygen-neon-magnesium (ONeMg) WDs (Schwarz et al. 1997, 2001, 2007, 
Rudy et al. 2003). V705 Cas (Nova Cas 1993) was probably to date the best studied case showing 
such a phenomenon \citep{1995ApJ...448L.119G, 1996MNRAS.282.1049E, 2005MNRAS.360.1483E}. 

The recent application of optical interferometers to the study of nova eruptions, not only provide 
information on the diameter and the shape of the ejecta \citep{1993AJ....106.1118Q, 2007ApJ...669.1150L}, but 
also on the physical processes at work. The state of the art facility of the VLTI,  both in terms of infrastructure and management of the observations,  allows one to schedule fast evolving novae observations 
\citep{2006SPIE.6268E..19S}. A snapshot of the early phase of the outburst (t=5.5d) of the famous 
recurrent nova RS\,Oph was caught by AMBER, the near-IR VLTI recombiner, allowing to reveal the complex 
geometry and kinematics of the ejection studied at a spectral resolution of 1500 \citep{2007A&A...464..119C}. 
\citet{2006ApJ...647L.127M} and \citet{2007ApJ...658..520L} reported on the monitoring of the RS Oph outburst 
using several near-IR optical interferometers during two months following the 2006 outburst. These 
observations allowed them to follow the expansion of the source and give some estimates of the 
free-free emission from the nova wind. RS\,Oph was also observed in the N band with the Keck 
Interferometer Nuller $\sim$3.8 days after discovery that provided evidence that some nebular lines 
and hot dust were present outside the obscured area, i.e. {\it before} the arrival of the blast wave 
of the outburst \citep{barry2008}.

V1280 Sco was discovered in outburst by \citet{2007IAUC.8803....1Y} on February 04.86 2007, about 12 
days before reaching its maximum in visual light (m$_V\sim$4). This climb to maximum is 
exceptionally slow, and 
the nova  also exhibited an unusual characteristic by forming dust as soon as two weeks after 
maximum. The first description of its infrared spectrum can be found in \cite{2007IAUC.8809....1R}. 
\citet{2007CBET..864....1D} report near-infrared JHK spectroscopy of V1280 Sco,
obtained with the Mt. Abu 1.2-m telescope displaying several strong C I emission lines. Further 
spectra, obtained on Mar. 4.95 UT showed a sharply rising continuum in the J, H, and K bands, indicating 
dust formation in the nova ejecta, a few days after a clear change is seen in the slope of the visual light curve. 
 \begin{table*}
\begin{caption}
{Observing log reporting the interferometric observations used in the paper.  In the column UTC Date are reported the date and hour of observation in the format year-month-day-T-hour. In the column Base are reported the different telescopes and stations used: the 8.2m fixed unit telescopes are labeled with U and the number of the telescope; the stations of the movable auxiliary telescopes are labeled with the letters A, G, H and K and a number. The reader can have an overview of the array in the following link: http://www.eso.org/observing/etc/doc/vlti/baseline/vltisations.html.}\label{table-log}
\end{caption}
\begin{tabular}{lllccccccc}\hline\hline 
Julian Day & 2007 UTC Date   & Day$^1$  & Instrument & Magnitude & Telescope & \multicolumn{2}{c}{Projected baseline} & Airmass& Closest \\
& & && & stations & Length  & PA  && Calibrator$^2$ \\
& &&& &&[meter] & [degrees] \\
\hline
2454160.4 & 2007-02-28T09 & 23 & AMBER(K)& 3.8 (K) & G1 - H0 & 71 & 175 & 1.14& HD151680\\
2454173.3 & 2007-03-13T08 & 36  & MIDI(N) &  1 (N) & G0 - K0 & 60.2 & 53 & 1.10 & HD150798\\
2454181.3  & 2007-03-22T07 & 45 & AMBER(K) & 4.2 (K) & G0 - H0 & 29.0 & 46 & 1.21 & HD149447\\
2454182.4  & 2007-03-23T08 & 46 & MIDI(N) &  0.3 (N) & A0 - G0 & 62.7 & 61 &  1.04 &HD150798\\
2454182.4  & 2007-03-23T09 & 46 & MIDI(N) &  0.3 (N) & A0 - G0 & 63.9 & 67 & 1.01 & HD150798\\
2454226.7  & 2007-05-06T05 & 90 &MIDI(N) & -0.8 (N) & U3 - U4 & 60.1 & 102 & 1.04 &HD163376 \\
2454226.9  & 2007-05-06T09 & 90 &MIDI(N) & -0.8 (N) & U3 - U4 & 57.9 & 132 & 1.18 &HD163376\\
2454227.8  & 2007-05-07T08 & 91 &MIDI(N) & -0.8 (N) & U3 - U4 & 59.3 & 128 & 1.11&HD163376\\
2454246.5  & 2007-05-26T01 & 110 & MIDI(N) & -1.5 (N) & U3 - U4 & 35.0 & 75 & 1.69 &HD163376\\
2454281.5  & 2007-06-30T00 & 145 & MIDI(N) & -1.1 (N) & U3 - U4 & 52.8 & 93  & 1.17 &HD123139\\
2454281.7  & 2007-06-30T05 & 145 & MIDI(N) & -1.1 (N) & U3 - U4 & 58.7 & 129 & 1.15 &HD163376\\
2454281.8  & 2007-06-30T07 & 145 & MIDI(N) & -1.1 (N) & U3 - U4 & 49.5 & 166 & 2.07 &HD177716\\

\hline
\multicolumn{7}{l}{\small $^1$ From discovery, Feb. 4.85 UT. JD=2454136.85}\\
\multicolumn{10}{l}{\small $^2$ HD151680 (K2.5III, 5.83$\pm$0.06), HD150798 (K2III, 8.76$\pm$0.12), HD149447 (K6III, 4.55$\pm$0.05), HD163376 (M0III, 3.79$\pm$0.12),}\\
\multicolumn{10}{l}{HD123139 (K0III, 5.12$\pm$0.02), HD177716 (K1III, 3.72$\pm$0.07)}\\
\end{tabular}
\end{table*}

We report on the first infrared interferometry observations ever recorded of a dust forming nova, 
providing a unique set of spectra and visibilities obtained during the first 145 days of the eruption 
of V1280 Sco. It was initially planned to exclusively use  the VLTI/AMBER near-IR combiner, but as  
the nova continued  producing dust at a high rate, most of the observations were shifted to the 
VLTI/MIDI mid-IR combiner. We were able to follow the expansion of the dust shell for more than one 
hundred days and to detect the appearance of a new shell. This allows us to constrain simultaneously 
the extension and content of the newly formed dust by means of simple geometric models and also by using 
the public domain {\tt DUSTY} code \citep{{1999astro.ph.10475I},1997MNRAS.287..799I}. It must however be emphasized that the parameter space characterizing the dust shells was not extensively covered and that the parameters proposed in the present article have to be considered as indicative only pending  further detailed investigations.
The outline of the paper is as follows: in Section 2 we present the analysis of the interferometric and 
spectroscopic data, in Section 3 we present a semi-quantitative interpretation of the event by means of 
the light curve and associated classical relations, in Section 4 we analyze the interferometric data using 
simple geometrical models and in Section 5 an attempt is made to investigate further the shell parameters 
using the {\tt DUSTY} code. Finally, we discuss in Section 6 some mechanisms for dust formation around novae 
that are able to account for observed the multiple shell behavior; the conclusions are presented 
in Section 7.

\section{Observations}
 
\subsection{A complex light curve}
\begin{figure}
 \centering
\includegraphics[height=9.cm]{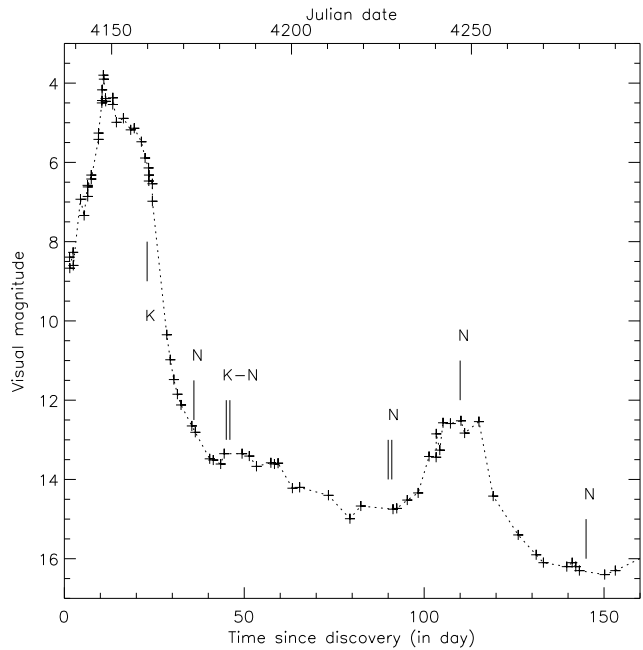}
\hfill
 \caption[]{The visual light curve of V1280 sco from AFOEV data with  the dates of the present VLTI observations superposed on it. The time for the light curve to drop by 2 magnitudes, t$_2$,  seems to occur after the beginning  of dust formation which is suggested by the sharp change in slope of the light curve. 
\label{fig:lightCurve}}
\end{figure}

Once a nova has been discovered, its light curve provided mostly by amateur observations, is the first 
measurement used to infer about the nature of the event. The light curve of V1280\,Sco, characterized by some 
interesting features, is shown in Fig.\ref{fig:lightCurve}. This curve was extracted from the AFOEV database, operated 
at CDS\footnote{http://cdsweb.u-strasbg.fr/cgi-bin/afoevList?sco/v1280}.
The first notable feature is the slow rise to visual maximum lasting about 12 days since the discovery
was reported.
Very few novae have exhibited such long
premaximum behavior which is generally found only in slow novae. The maximum of visual light on the 16th of February is characterized by a sharp rise from the 5th to the maximum magnitude m$_{\rm v}\sim$ 3.8. This is followed by a smoother decline lasting less than 10 days, interrupted abruptly by the dust formation event. The source  faded over the next few days from  7th to the 12th magnitude, then continued to fade at a lower rate to reach and stabilize at 15th magnitude between t=80 and t=90 days (2007-04-27/2007-05-06). A second maximum occurred, peaking near t=104d (2007-05-20) with a magnitude m$_{\rm V}$ = 12.5 followed again by a sharp decline to the 16th magnitude. 
The sequence of our observations is as follows. The first AMBER observations were performed on the 28th of February, while the nova had not yet formed dust. The next observations, originally planned with AMBER, were carried out on the 13th of March with MIDI as it appeared that dust was forming at a high rate. The first MIDI observations were successfully performed with the 1.8m Auxiliary Telescopes (ATs) on the 13th of March. These observations were difficult to reduce because the N band flux of V1280~Sco of 15-20Jy was close to the MIDI sensitivity limit. Only the visibility measurement performed at the lower airmass ($\sim$1.1) could be scientifically used, the two other measurements having error bars that are too large. On the 22nd of March, AMBER observations with 2 ATs were performed followed by MIDI observations on 23rd March. In May, the visual flux dropped below the sensitivity limit of m$_{\rm V}$ = 13.5, which is required to acquire the source on the ATs.
Therefore, the observations were then conducted using the 8.2m Unit Telescopes (UTs) although the source was at this epoch even brighter in the N band, rising to m$_{\rm N}$ = -1.5 on the 26th of May. We were fortunate enough to get observations at some crucial moments during the evolution of the nova. The first K band observations were made at t=23d (2007-02-28) just before the dust formation event, while the subsequent observations cover the evolution from t=36d (2007-03-13) to the minimum light at t=90d (2007-05-06). One observation was recorded near the top of the second maximum at t=110d (2007-05-26) while the last observations at t=145d (2007-06-30) cover the second deep minimum.

\begin{figure*}
 \centering
\includegraphics[width=16.5cm]{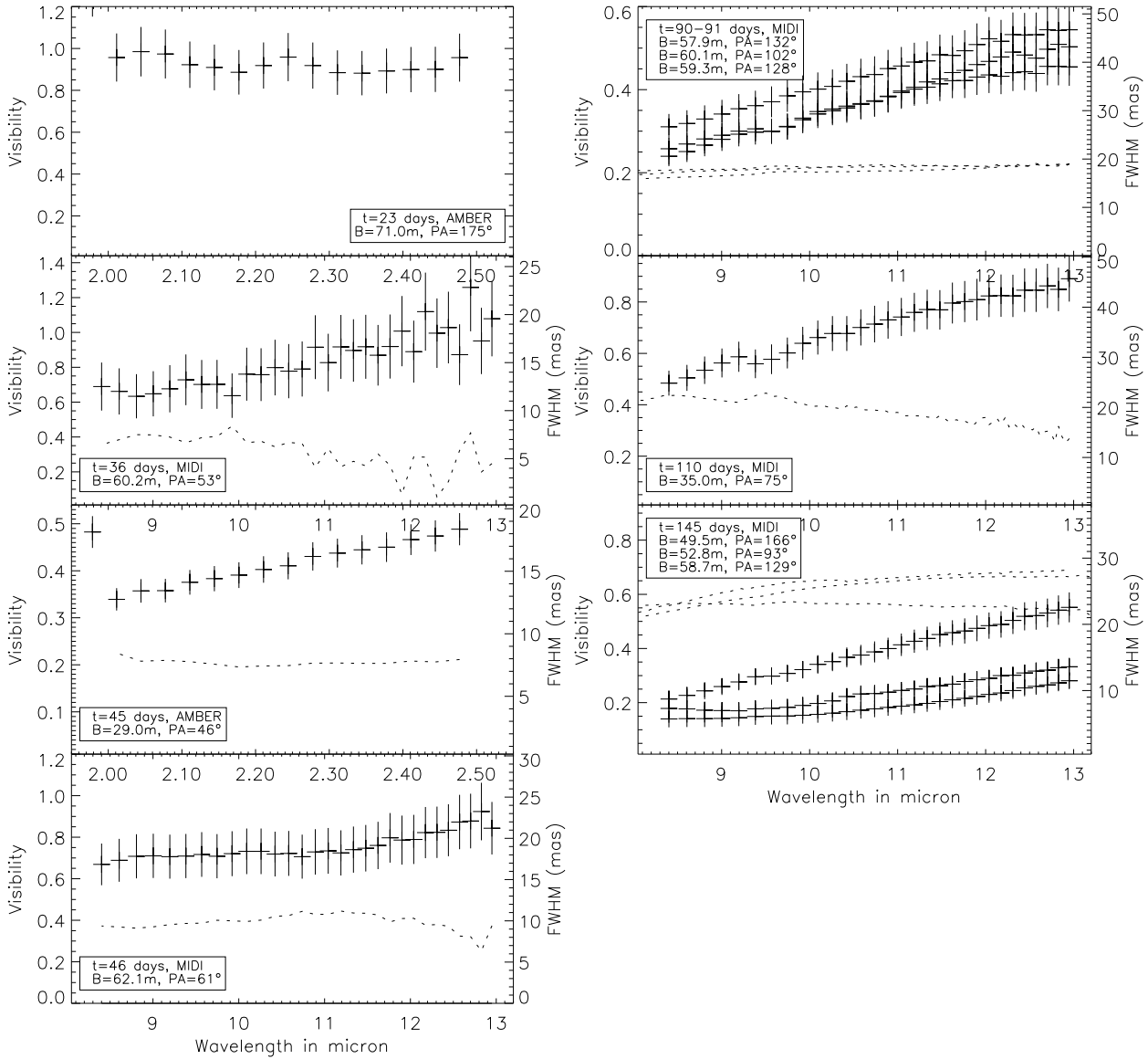}
\hfill
 \caption[]{Full data set of visibilities recorded with MIDI and AMBER, from the earliest dates (top, left) to the latest ones (bottom, right). The visibilities are indicated by the crosses (left vertical axis), and  estimates of the size of the source, using Gaussian models is shown in dotted lines (right vertical axis).  
\label{fig:vis}}
\end{figure*}

The sequence of VLTI observations is described in Table \ref{table-log}. We used the VLT Unit Telescopes (8.2m, UTs), whose stations are labeled U2, U3 or U4 in the table and the Auxiliary Telescopes (1.8m, ATs), which were located at stations A0, G0, G1, H0, K0. The spectrally dispersed visibilities are shown in Fig.\ref{fig:vis}
 
\subsection{AMBER data}
The AMBER data were recorded using the combination of 2 telescopes only, as the three telescope 
mode was difficult to organize and operate under conditions with a sense of urgency
needing observations to be planned at short notice \footnote{For instance, without 
having  accurately set the Optical Path Delay model for the new baselines by several observations of the calibrators.}. For the same reasons, the data were secured using the lowest spectral resolution 
($\lambda / \delta \lambda$=35). The AMBER data were reduced using the
amdlib 
software developed by the AMBER consortium following the scheme described by Tatulli et al. (2007). The
squared visibility estimator is computed from the basic observables
coming from the used algorithm: the coherent flux (i.e. complex
visibilities obtained frame by frame multiplied by the flux) and the estimated
fluxes from each telescope.

These low resolution data were reduced applying the 'standard' AMBER data reduction
recipes i.e. selecting files where the optical path difference
measurement does not exceed half the coherence length ($\lambda^2 /
\delta \lambda$) and then applying frame selection, rejecting 80\% of
the 'bad' frames, according to the fringe contrast SNR measurement
(as explained in Millour et al. 2007). Only one calibration star was observed for each night, 
HD\,151680 (K3III, 5.93$\pm$0.06mas) and HD\,149447 (K6III, 4.55$\pm$0.05mas) for t=23d and t=45d respectively. In this context, accounting for the error bars and systematics is not straightforward. 
Therefore we consider a conservative value of 10\% for the relative accuracy on $V^2$, much larger 
than the one we can get from internal dispersion of the squared visibilities (4\% for the first 
dataset and 2.5\% for the second). On  other closeby nights, the transfer function scatter is of 
the order of 10\% rms (i.e. 0.1 for a transfer function of 1 and 0.05 for a transfer function of 0.5).

The calibrated visibility of V1280\,Sco on the 28 Feb. is very close to 1, meaning that the object is unresolved within the large error bar estimates. The same comment applies to the night
of 22 Mar. 2007, but as the calibration star is smaller and the
visibility of V1280\,Sco is also smaller, the resulting error bar and
systematics are less prominent.

We tried to estimate whether the first K band AMBER data were exhibiting some evidence of a rising dust continuum by performing a (crude) flux calibration using the interferometric calibrators as 
flux standards. The 28 Feb spectrum shows clusters of emission lines, not resolved with the low spectral resolution mode, and a decreasing continuum towards longer wavelength. The spectrum taken 22 days after discovery is much smoother and a significant increase of flux toward long wavelengths is observed.

\subsection{MIDI data}
The VLTI/MIDI interferometer operates like a classical Michelson interferometer to combine the mid-IR light (N band, 7.5-13.5$\mu$m) from two UTs or two ATs. We used a typical MIDI observing sequence, as described in Ratzka et al. (2007), that is briefly summarized here. Chopped acquisition images are recorded (f=2Hz, 2000 frames, 4~ms per frame, 98mas per pixel) for the fine acquisition of the target. The beam combiner and the dispersion unit are then introduced in the optical path and the interference pattern scanned by means of a temporal modulation of path by a reflective device mounted on a piezo motor. The lower spectral resolution provided by a prism was mostly used providing a spectral dispersion $\lambda / \delta \lambda$ of about 30, but some data were recorded with the high spectral mode provided by a grism ($\lambda / \delta \lambda$=230) although we note that all the data recorded were spectrally featureless. The data reduction softwares\footnote{\tt{http://www.mpia-hd.mpg.de/MIDISOFT/, http://www.strw.leidenuniv.nl/$\sim$nevec/MIDI/}} MIA and EWS were used to reduce the spectra and visibilities.

\begin{figure}
 \centering
\includegraphics[width=9.cm]{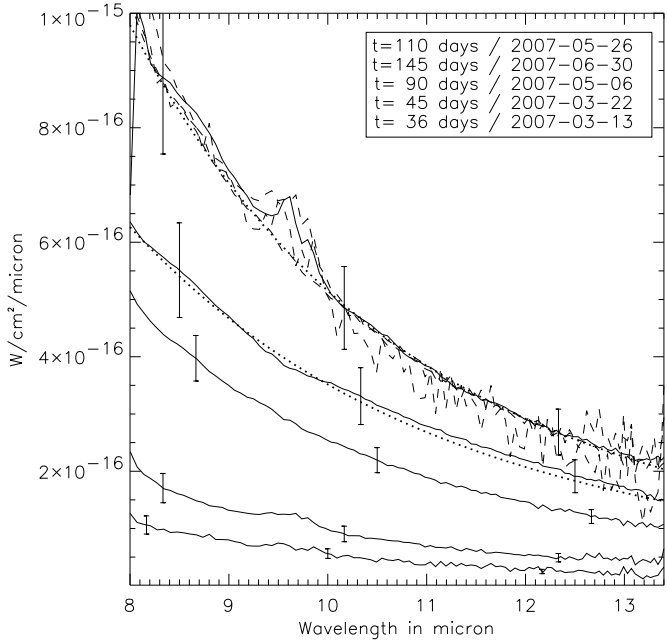}
\hfill
 \caption[]{From bottom to top, flux calibrated mid-IR spectra recorded on days 36, 45, 90, 145 and 110. The dashed lines superimposed on the maximum flux recorded at day 110 are the fluxes at t=36, 45 and 90 days after scaling. The slopes of the four spectra are similar within error bars indicating an invariant temperature for the dust continuum. The t=110 days curve is well fitted by a black-body with T=920K while that for t=145 days exhibits a slightly different slope which is  well fitted by a black-body with T=760K. The two fits are indicated in dotted lines.
\label{fig:MIDI_flux}}
\end{figure}

Most of the MIDI observations of V1280\,Sco were performed in the so-called HIGH\_SENS mode, meaning that the photometry of each telescope is recorded subsequently to the fringes. The errors, including the internal ones and the ones from the calibrator diameter uncertainty, range from 20\% for the first dataset, at the limit of sensitivity of AT telescopes, 10-15\% for the AT observations and about 7\%-15\% for the UT observations. The accuracy of the absolute flux calibration is in the 10-15\% interval, but the slopes are accurate at the percent level. The fluxes are shown in Fig.\ref{fig:MIDI_flux}, together with the fits to the data using blackbody curves indicating their effective temperature.

Accurate differential phase curves could be extracted with EWS from the good quality data recorded on the 6th of May and later on. The curves from V1280\,Sco are smooth and dominated by atmospheric residuals of the same order as that arising from the calibrators. The corrected differential phase is typically 0\deg$\pm$5, for each day of observation. These measurements are indicative that the source does not present a strong departure from point symmetry, but does not provide any information on possible departures from spherical geometry (e.g. bipolar geometry). The EWS software removes a large part of the low frequency phase signal, but a strong phase signal from dusty disks can still be detected unambiguously \citep{2007A&A...467.1093D}.

\subsection{Near-Infrared spectroscopy}

\begin{figure}
 \centering
\includegraphics[width=8.5cm]{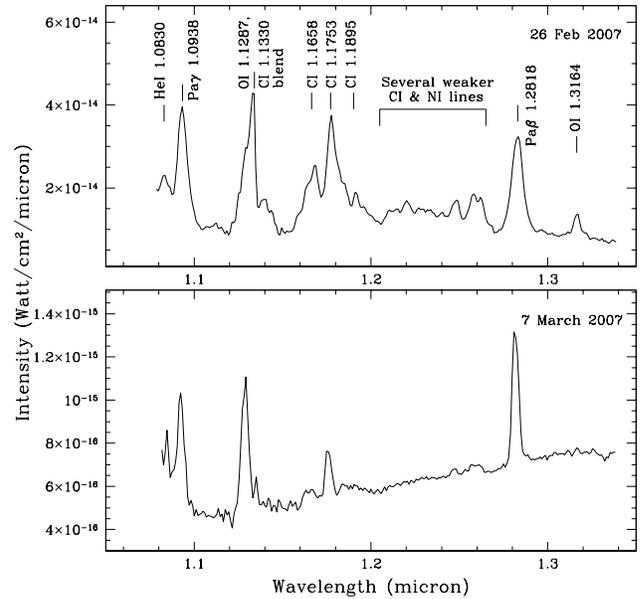}
\hfill
 \caption[]{J band spectrum of V1280 Sco on February 26.97 (t=22d, top panel) and March
7.95, 2007 (t=38d, bottom panel) are shown. The prominent emission lines are
marked. The formation of dust between the two epochs is evidenced
clearly from the rising of continuum towards longer wavelengths in the
later spectrum of 7 March 07, while no dust signature is visible in the first spectrum.
\label{fig:IRspectra}}
\end{figure}

Since its outburst in February 2007, near-infrared JHK observations of
V1280 Sco were made on a regular basis from the 1.2m telescope at
the Mt.  Abu Observatory, India. A detailed presentation of these
results is made elsewhere \citep{das2008} though an early set of spectra, taken soon after
the outburst, is described in Das et al. (2007a) and Das et al. (2007b).
In the present work, we show in Fig.\ref{fig:IRspectra}, representative J band spectra
of V1280 Sco taken on 26 February 2007 and 7 March 2007. The
observations were obtained with the Near Infrared Imager/Spectrometer
with a 256x256 HgCdTe NICMOS3 array at a resolution of approximately 10
Angstroms/pixel. The J band spectra are typical of a
classical nova soon after outburst, showing prominent lines of HI, HeI, OI
and CI. The presence of strong neutral CI lines is invoked
subsequently (while modeling the dust properties) to be indicative
that carbon is the likely principal constituent of the dust
that formed around V1280 Sco. The significant change seen in the slope
of the continuum, between the two epochs of observations in Fig.\ref{fig:IRspectra},
is also indicative of the formation of dust around this time (this
is also manifested by a sharp decline in the optical light curve at
approximately this epoch). The later spectrum of 7 March shows a
steepening of the continuum towards longer wavelengths i.e. the
building up of an infrared excess which is expected during the onset of
dust formation. 
Several K band spectra, recorded between the 18th of February and the 8th of June, are shown in \citet{das2008}. The 27th of February spectrum exhibits a strong Br$\gamma$ lines peaking at twice the continuum level, and few weaker broad lines. The Br$\gamma$ line is maximum the 1rst of March, peaking at 3.4 above the continuum and then decreases markedly. The spectrum recorded the 13th of March is almost featureless, dominated by the dust continuum. The first AMBER data were secured the 28th of February, before the starting of the dust formation. The second, secured the 13th of March, probes essentially the dusty shell.

\begin{table*}[]
	\centering
		\begin{tabular}{llll|cccc|cccc}
			\hline
			\hline
			Day & band &\multicolumn{2}{c|}{Projected baseline}&\multicolumn{4}{c|}{Uniform Disk Diameter (mas) }&\multicolumn{4}{c}{Gaussian FWHM (mas)}\\
&&B (m) & PA ($^\circ$) & 2.2$\mu$m & 8$\mu$m&10$\mu$m&13$\mu$m&2.2$\mu$m &8$\mu$m&10$\mu$m&13$\mu$m\\
			\hline
23 &K&71&175&$\leq1.5$&&&&$\leq1.0$\\	
36&N&60.2&53&&8$\pm$3.0&10$\pm$3.0&7$\pm$4.0&&5$\pm$2.0&7$\pm$2.0&5$\pm$3.0\\
45&K&29&46&12.5$\pm2$&&&&7.5$\pm$1.\\
46&N&62.7&61&&13$\pm$1.5 &13$\pm$1.5& 12$\pm$2.5 &&8$\pm$4.0 &8$\pm$4.0 &7$\pm$4.0\\
90&N&60.1&102&&25$\pm$0.5&28$\pm$0.5&30$\pm$1.0&&17$\pm$1.0 &18$\pm$1.0& 19$\pm$1.0\\
90&N&57.9&132&&27$\pm$1.0 &28$\pm$1.0&30$\pm$1.0&&18$\pm$2.0 &19$\pm$2.5& 19$\pm$3.\\
91&N&59.3&128&&26$\pm$1.5&30$\pm$1.5&32$\pm$2.0&&18$\pm$1.5 &19$\pm$1.5& 21$\pm$2.\\
110&N&35.0&75&&35$\pm$3.0&34$\pm$3.0&32$\pm$3.0&&22$\pm$2.0&20$\pm$2.0&15$\pm$2.5\\
110$^*$&N&35.0&75&&43$\pm$3.0&43$\pm$3.0&42$\pm$3.0&&27$\pm$2.0&27$\pm$2.0&27$\pm$2.0\\
145&N&52.8&93&&37$\pm$1.0&44$\pm$1.0&47$\pm$2.0&&22$\pm$1.5&26$\pm$1.5&28$\pm$1.0\\
145$^*$&N&52.8&93&&-&48$\pm$1.0&51$\pm$2.0&&$\ge$35&33$\pm$1.5&33$\pm$1.0\\
145$$&N&58.7&129&&36$\pm$1.5&40$\pm$1.5&45$\pm$2.0&&21$\pm$1.0&24$\pm$1.0&27$\pm$1.5\\		
145$^*$&N&58.7&129&&-&-&50$\pm$2.0&&$\ge$35&33$\pm$1.5&34$\pm$1.0\\		
145&N&49.5&166&&38$\pm$4.0&39$\pm$3.0&37$\pm$3.0&&23$\pm$1.0&23$\pm$1.0&22$\pm$1.5\\
145$^*$&N&49.5&166&&42$\pm$1.0&42$\pm$2.0&42$\pm$2.0&&30$\pm$1&27$\pm$1.0&26$\pm$1.5\\
			\hline
		\end{tabular}
		\caption{\label{tab:geom} Results from the conversion of the MIDI visibilities by means of geometric models. The quoted errors are at the 1$\sigma$ level. The lines indicated by an asterix, close to the data, present the extension of the  model wherein we  assume an additional unresolved source representing 10\% of the total flux at t=110d and 15\% at t=145d (keeping in mind that the total flux from the unresolved source is a lower estimate of the true flux of the appearing compact source). For further details please see the text. }
\end{table*}

\section{Semi-quantitative interpretation}

The interferometric observables in the infrared domain provide a new insight on the dust shells formed around the novae that complements the information from the flux calibrated spectra. The modeling of the dust shell involves many parameters, representing the variable temperature and brightness of the source, the nature of the dust and the density law in the dust forming region -  the visibilities {\it per se} do not allow to suppress or resolve all the degeneracies. As a consequence, it becomes necessary to estimate a few  parameters using classical relations and velocity estimations in order to use them as a first guess in the fitting process.

\subsection{First estimation of the basic parameters: classical relations}

\cite{2007CBET..852....1M} measured a B magnitude of 4.1 and a V magnitude of 3.79 for V1280 Sco at the time of maximum visual light (the 16th of Feb.). The temperature a few days before maximum was close to a F 
type star \citep{2007IAUC.8803....1Y}, and a temperature of 5300K is also mentioned by \cite{2007IAUC.8809....1R}. We assume that the temperature at maximum was about T$_0$=7000K.
Assuming a B-V$\sim$0/0.3 (for an A or F spectral type), we can estimate E(B-V) to 0.3 or slightly less, 
and A$_V$ to 0.9-1.1. As a cross-check, the galactic extinction law from \cite{2006A&A...453..635M}, 
provides a K absorption coefficient A$_K$=0.107 to 0.127 between 1 and 8kpc, scaling to A$_V$=1.2-1.4 (assuming A$_V$/A$_K$=11).

One can use the relationship between absolute magnitude and rate of decline (the so-called MMRD relations, \citet{1995ApJ...452..704D}) for getting an estimate of the absolute magnitude and the distance of the source. The estimation of the $t_2$ and $t_3$ values i.e. the
time the nova took to fade by 2 and by 3 magnitudes respectively, gets complicated
because of the sudden drop in the light curves that interrupted the regular fading trend as a consequence of the fast formation of dust. If we ignore dust formation and extrapolate the light curve to see when 2 magnitudes of fading occurs, then 21 days appears as an appropriate estimate. For Av = 1.2, the MMRD relations yield D = 1.3 kpc for $t_2$=21 days and D = 1.7 kpc for $t_2$ = 13 days, the shortest estimate allowed by the dust forming event.
Also, two rates of decline can be extracted: one applicable from t=11d (maximum light) to t$\sim$16d, with a fast decline of 0.24mag/day, and a much slower from this point to t$\sim$21d, of less than 0.1 mag/day. The curve trend is then abruptly changed by the dust forming event.

In \cite{2007IAUC.8803....2N}, the FWHM of the H$\alpha$ emission on Feb. 5.87 UT is 400 km.s$^{-1}$, and 
its absorption minimum is blueshifted by 480 km.s$^{-1}$ from the emission peak. Low resolution spectra obtained on Feb. 12.88 and 14.83 UT \citep{2007IAUC.8807....2Y} show Balmer lines with clear P-Cyg profiles and an expansion velocity of about 500 km.s$^{-1}$. \cite{2007CBET..852....1M} report that, on an average, the displacement of the P-Cyg absorptions from the emission components is 595 km.s$^{-1}$. The Br$\gamma$ line also showed during the initial days a marked P Cygni profile centered at about 575 km.s$^{-1}$ \citep{das2008}. Thus, for the estimation of the size of the ejecta, we will adopt a velocity value of 500$\pm$100km.s$^{-1}$.

\section{Fitting the interferometric data with geometric models}
In this section, we interpret the visibility curves via simple geometric models assuming a uniformly distributed or Gaussian-like radially symmetric flux distribution. Table 2 displays the results of the fits at some chosen wavelengths.

\begin{figure}
 \centering
\includegraphics[width=9.5cm]{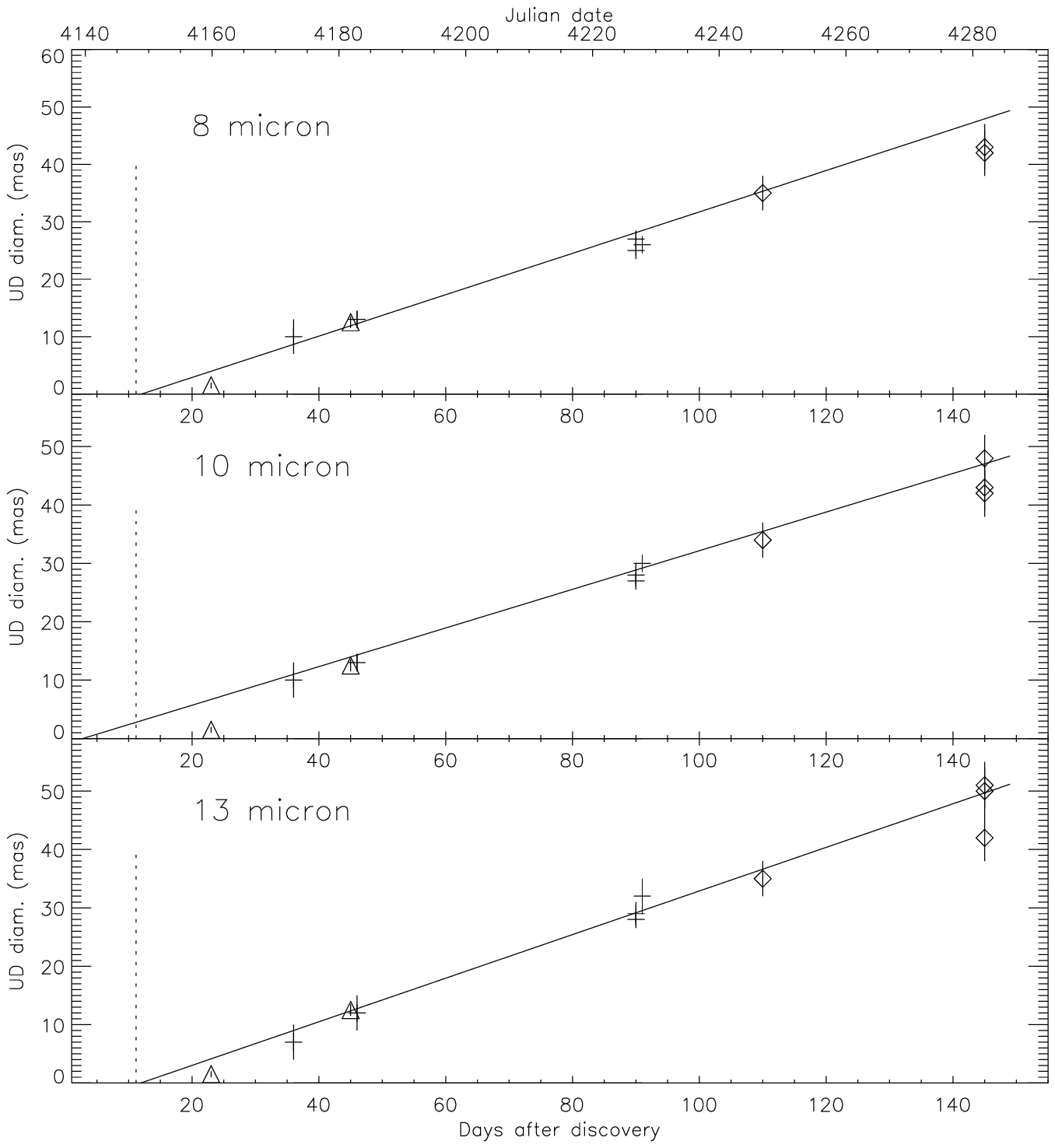}
\hfill
 \caption[]{Uniform disk (UD) diameters at 8$\mu$m (top), 10$\mu$m and 13$\mu$m (bottom) as a function of time, together with the best polynomial fits of a linear expansion model. The time of maximum light (t=11d) is shown as a dotted vertical line. The points are directly inferred from the visibilities using simple UD models (crosses), while the squares, representing the diameters measured after t=110d are corrected from the presence of a compact, unresolved source. Taking this correction into account (and its potentially large associated error), one can see that the first dust shell originates from material launched a few days before the maximum light. The AMBER measurements at t=23d and t=45d are shown as triangles (these points are not included in the fitting process).
\label{fig:expansion}}
\end{figure}

\begin{figure*}
 \centering
\includegraphics[width=9.cm,height=6.cm]{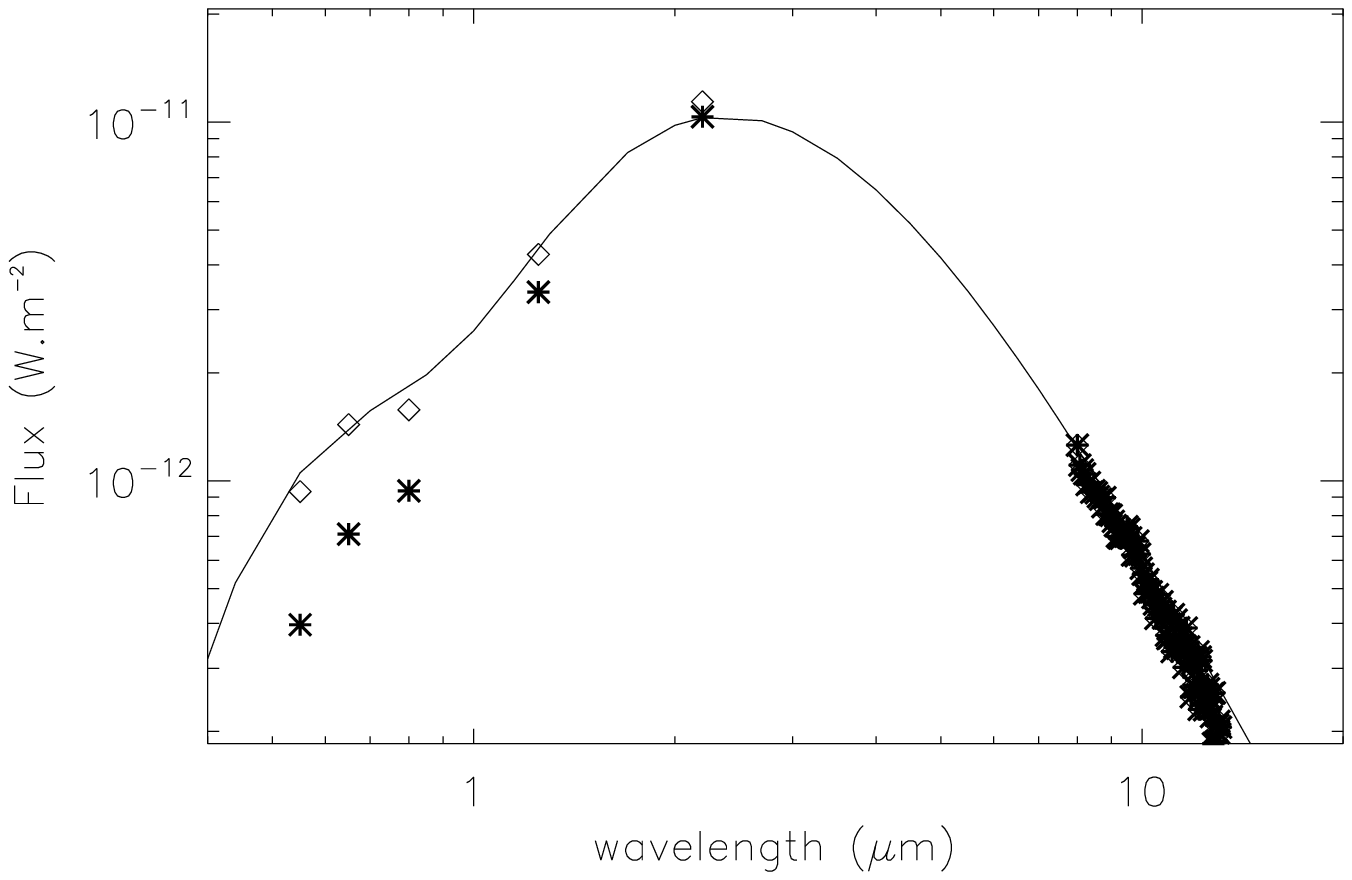}
\includegraphics[width=9.cm,height=6.cm]{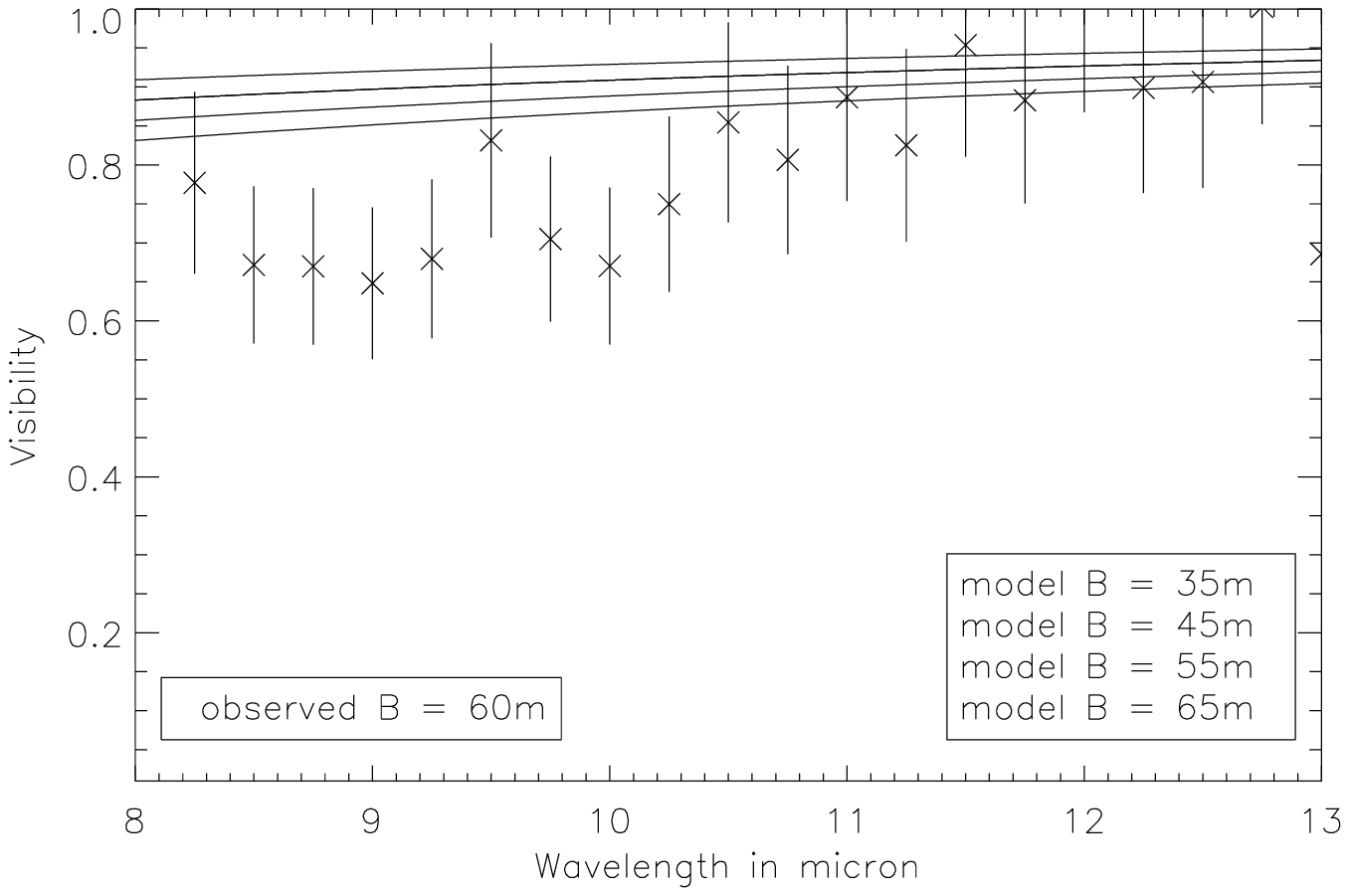}
\includegraphics[width=9.cm,height=6.cm]{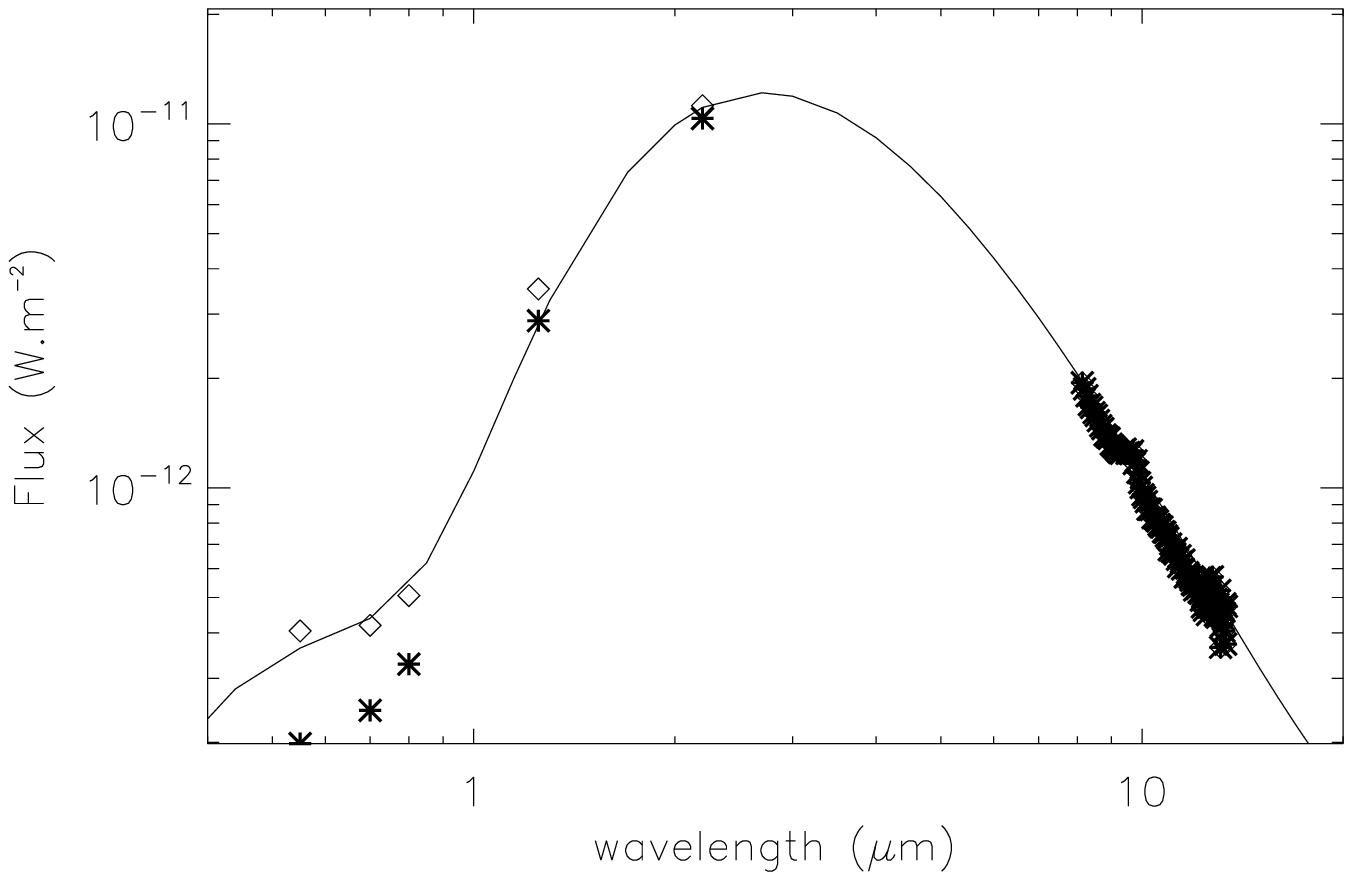}
\includegraphics[width=9.cm,height=6.cm]{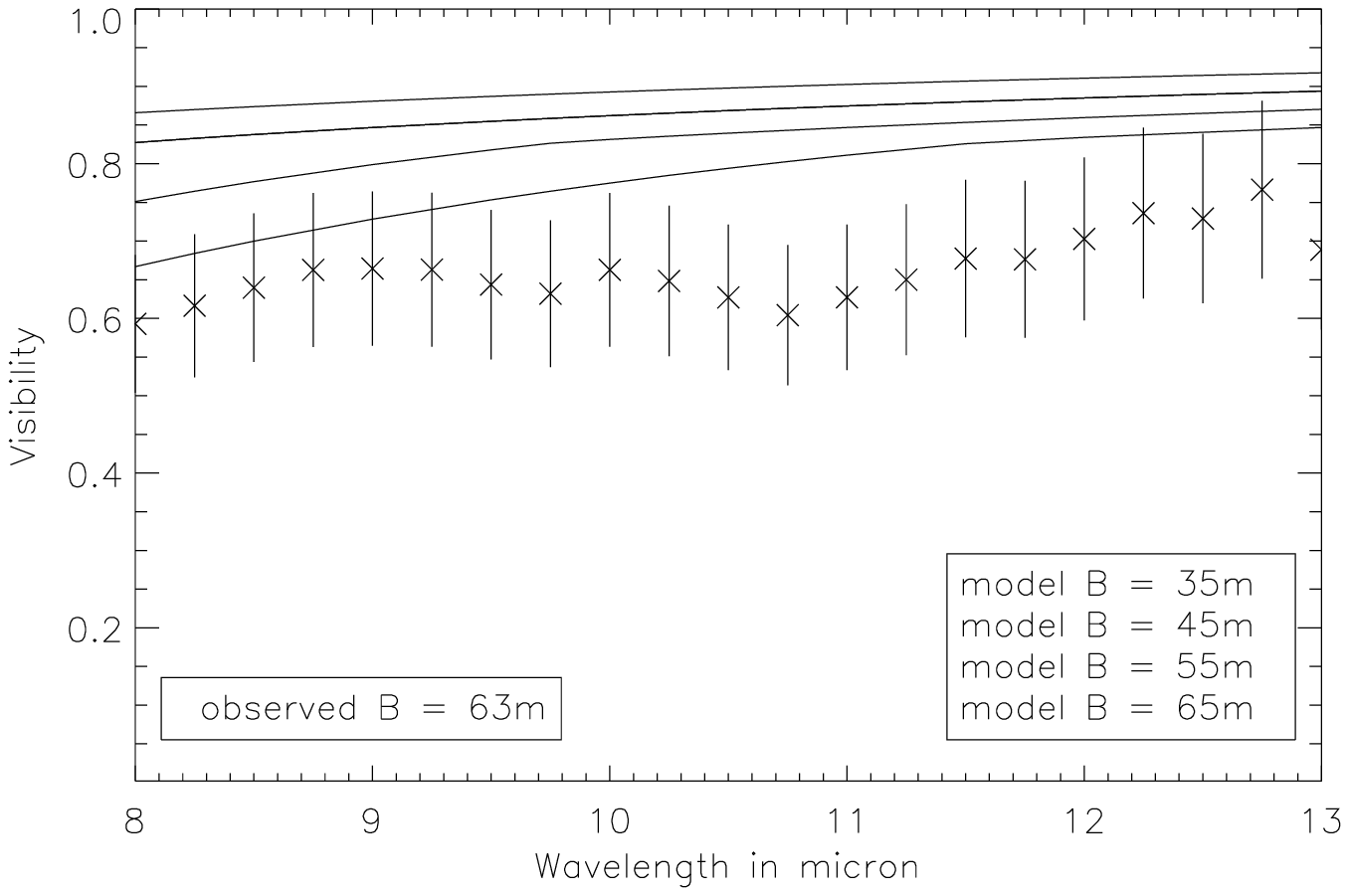}
\includegraphics[width=9.cm,height=6.cm]{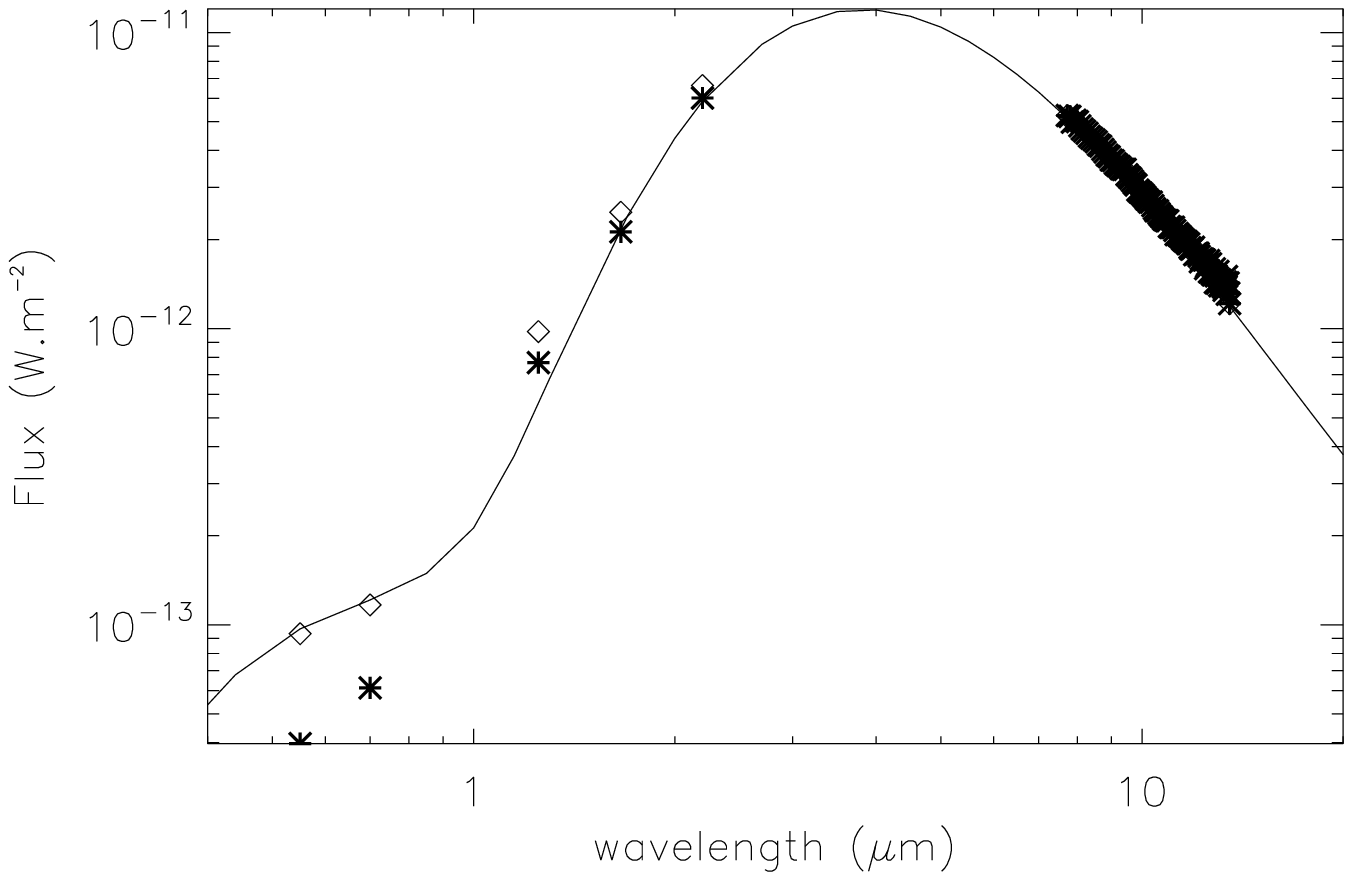}
\includegraphics[width=9.cm,height=6.cm]{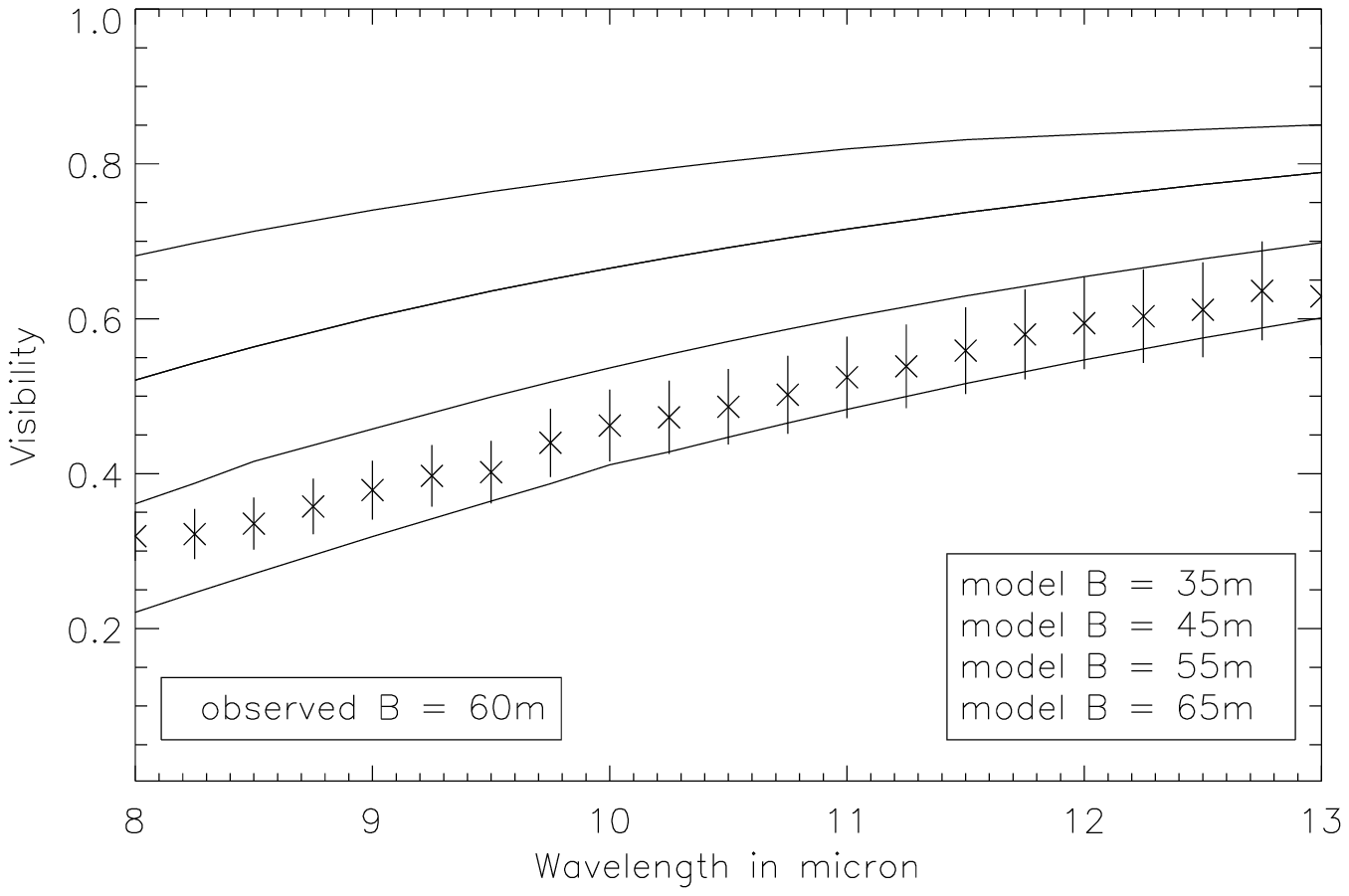}
\hfill
 \caption[]{From top to bottom, SEDs at left and visibilities on right recorded at day 36, 45 and 90 
compared to the {\tt DUSTY} models with parameters described in Table.\ref{table-model}. The SEDs are dereddened, assuming A$_V$=0.9, changing the observed fluxes (stars) to the corrected ones (squares).
\label{fig:models}}
\end{figure*}

\begin{figure*}
 \centering
\hfill

\includegraphics[width=9.cm,height=6.cm]{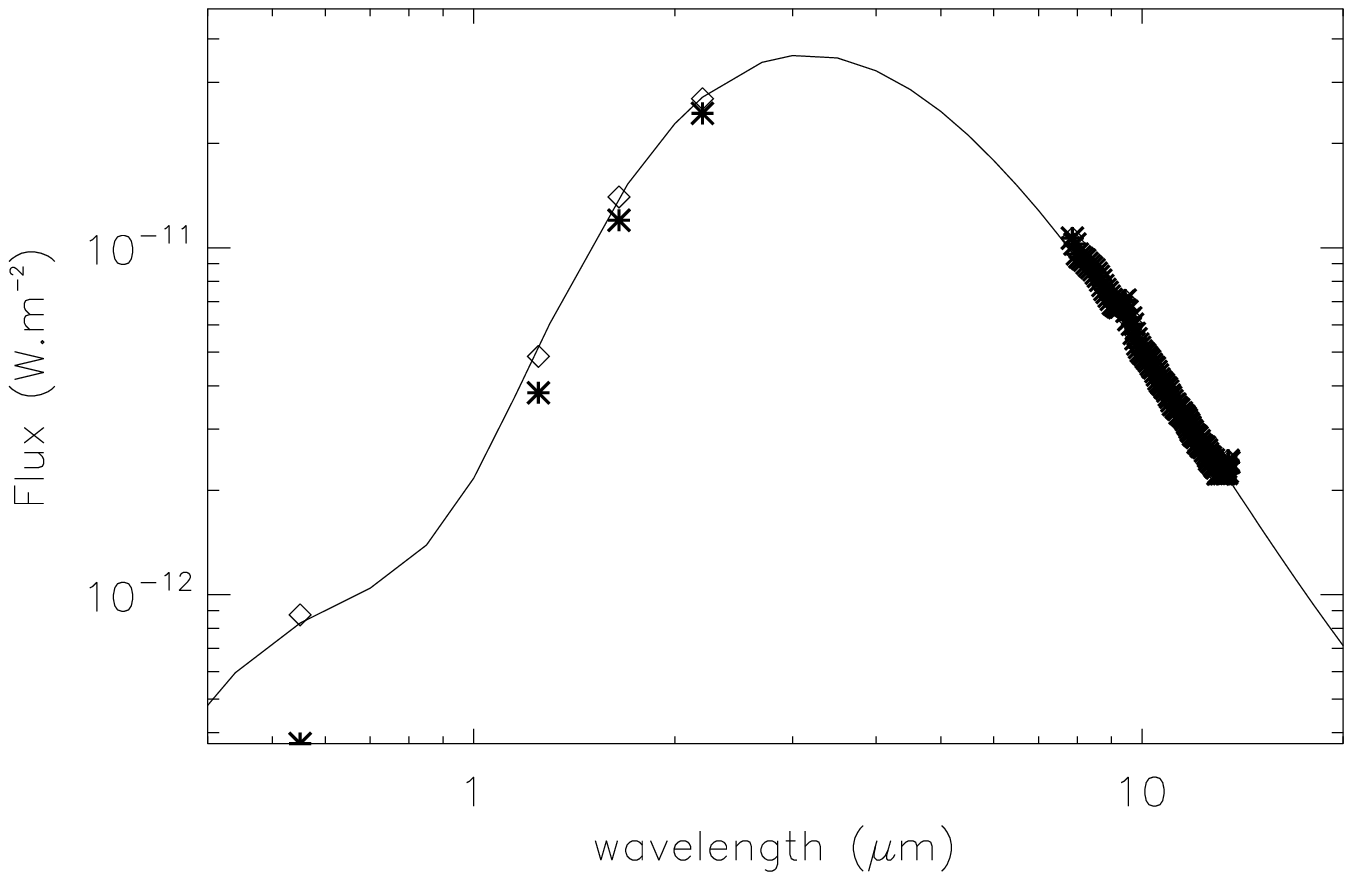}
\includegraphics[width=9.cm,height=6.cm]{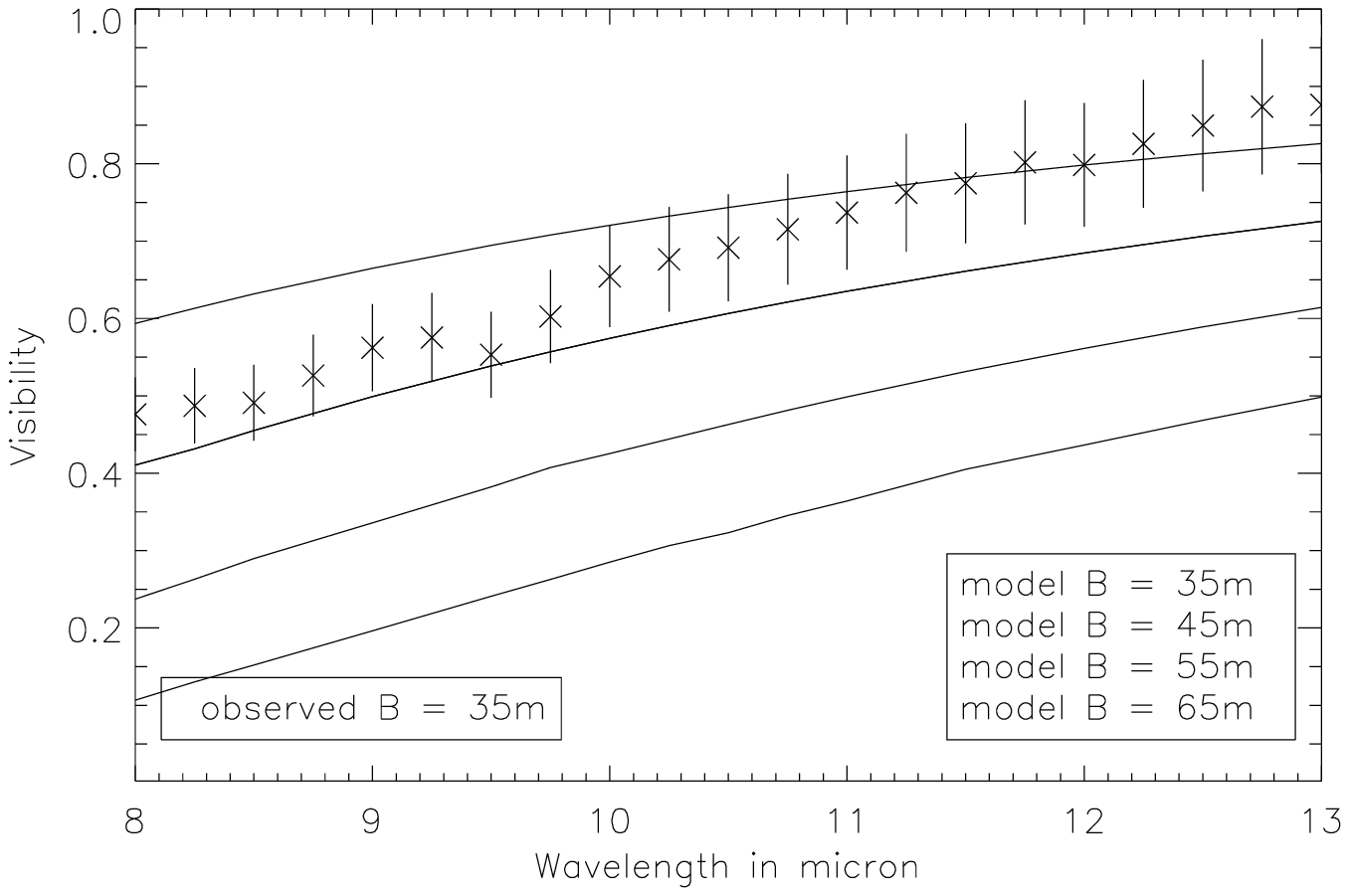}
\includegraphics[width=9.cm,height=6.cm]{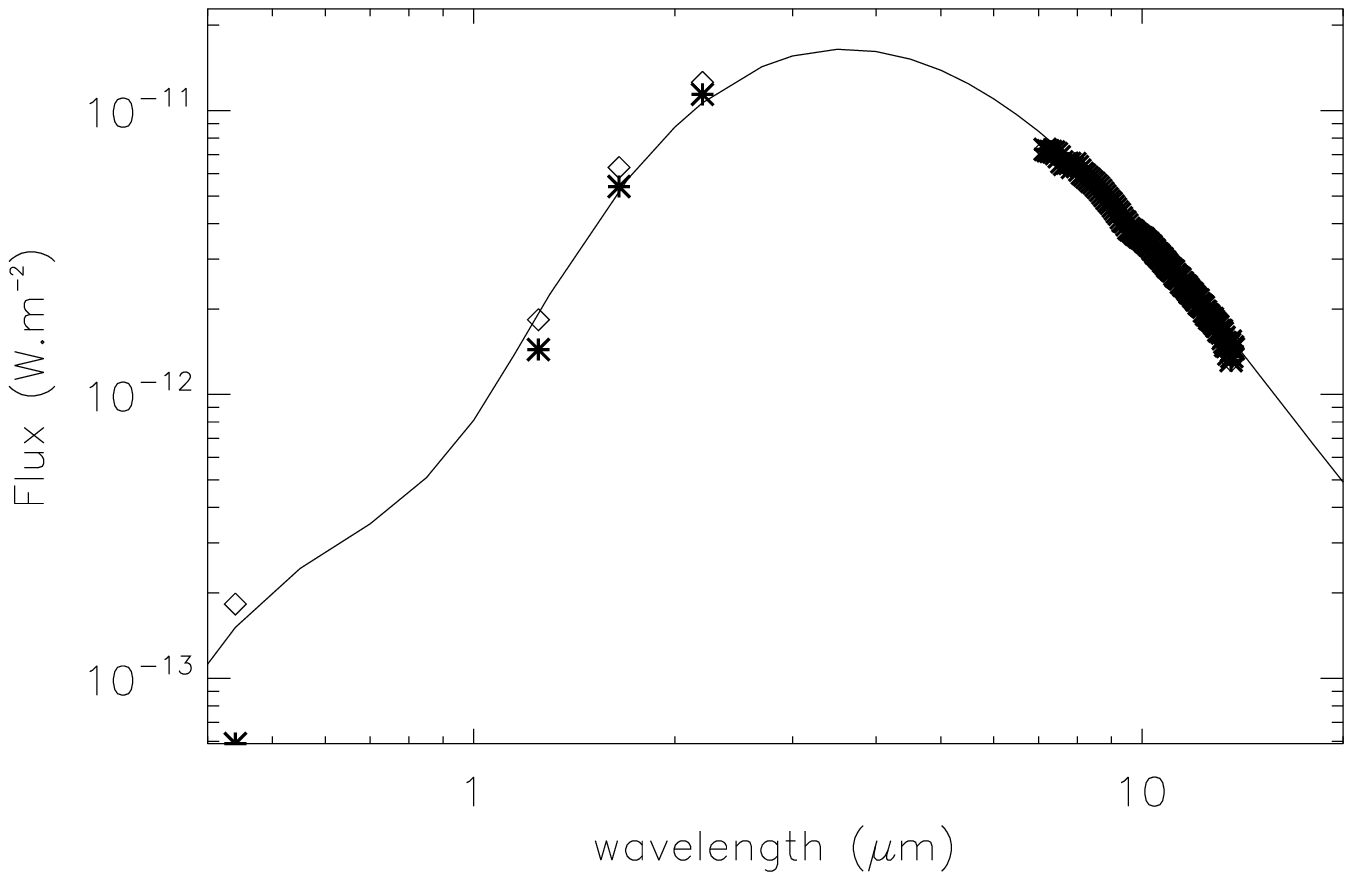}
\includegraphics[width=9.cm,height=6.cm]{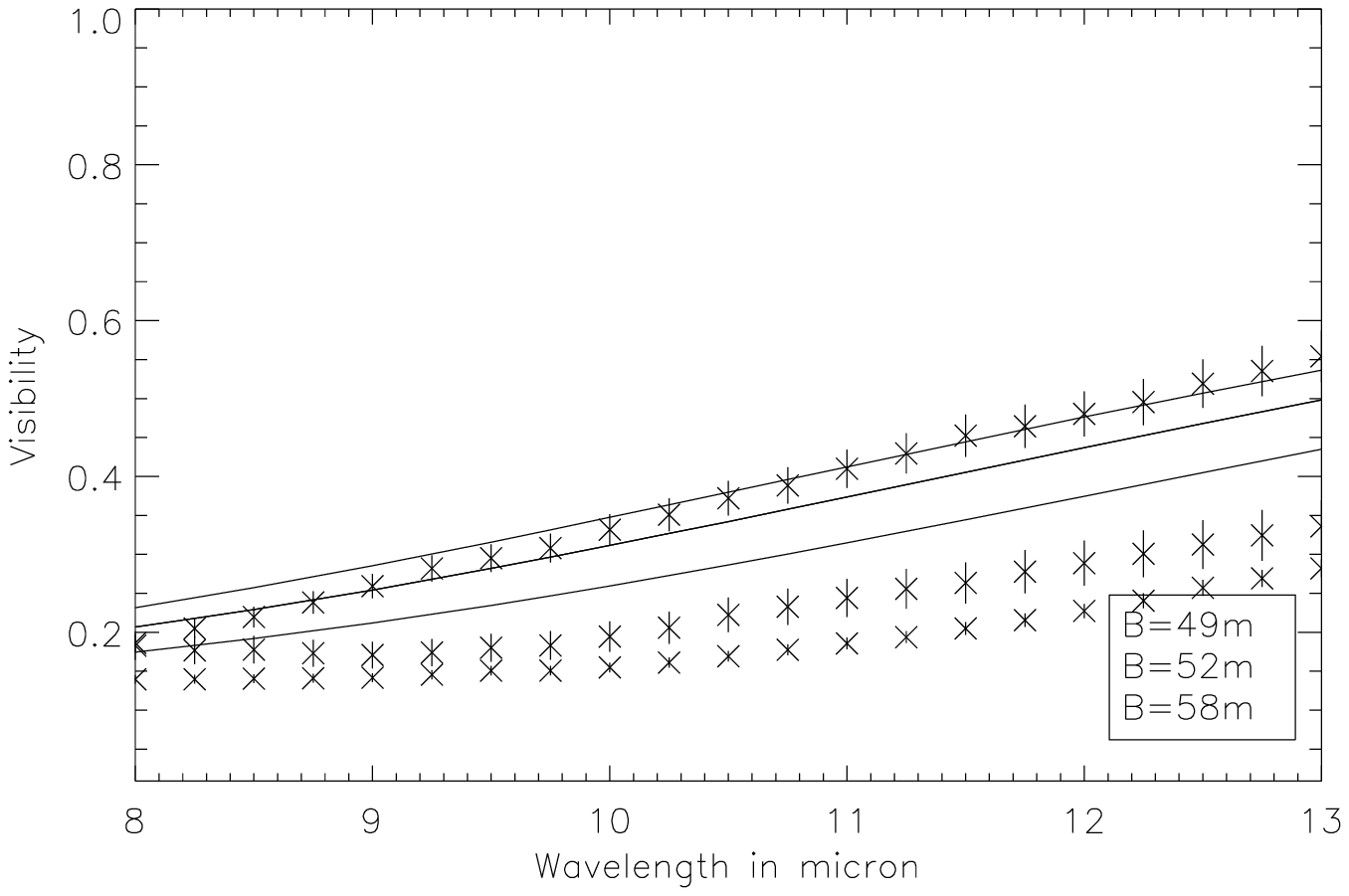}
 \includegraphics[width=9.cm,height=6.cm]{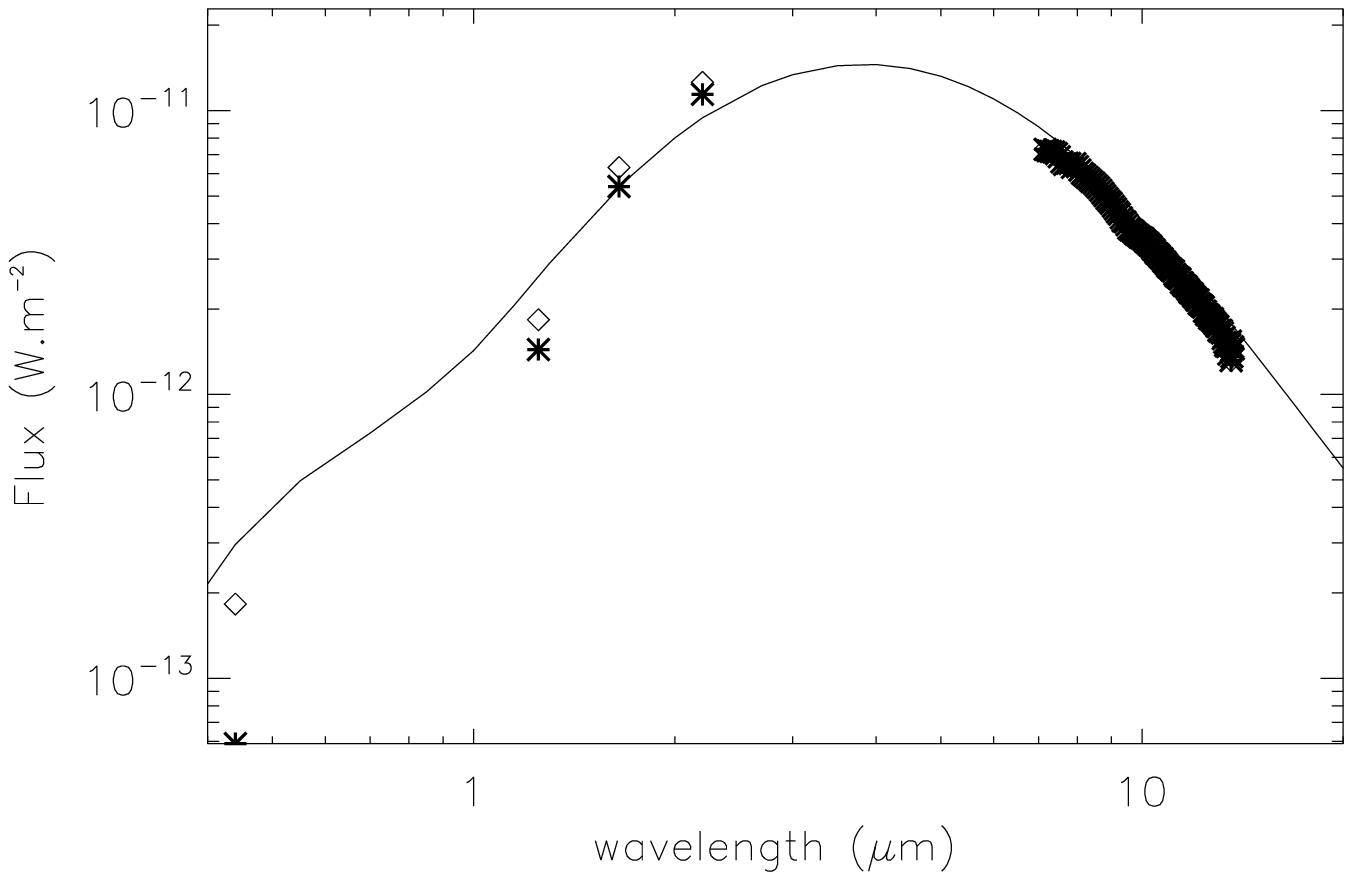}
\includegraphics[width=9.cm,height=6.cm]{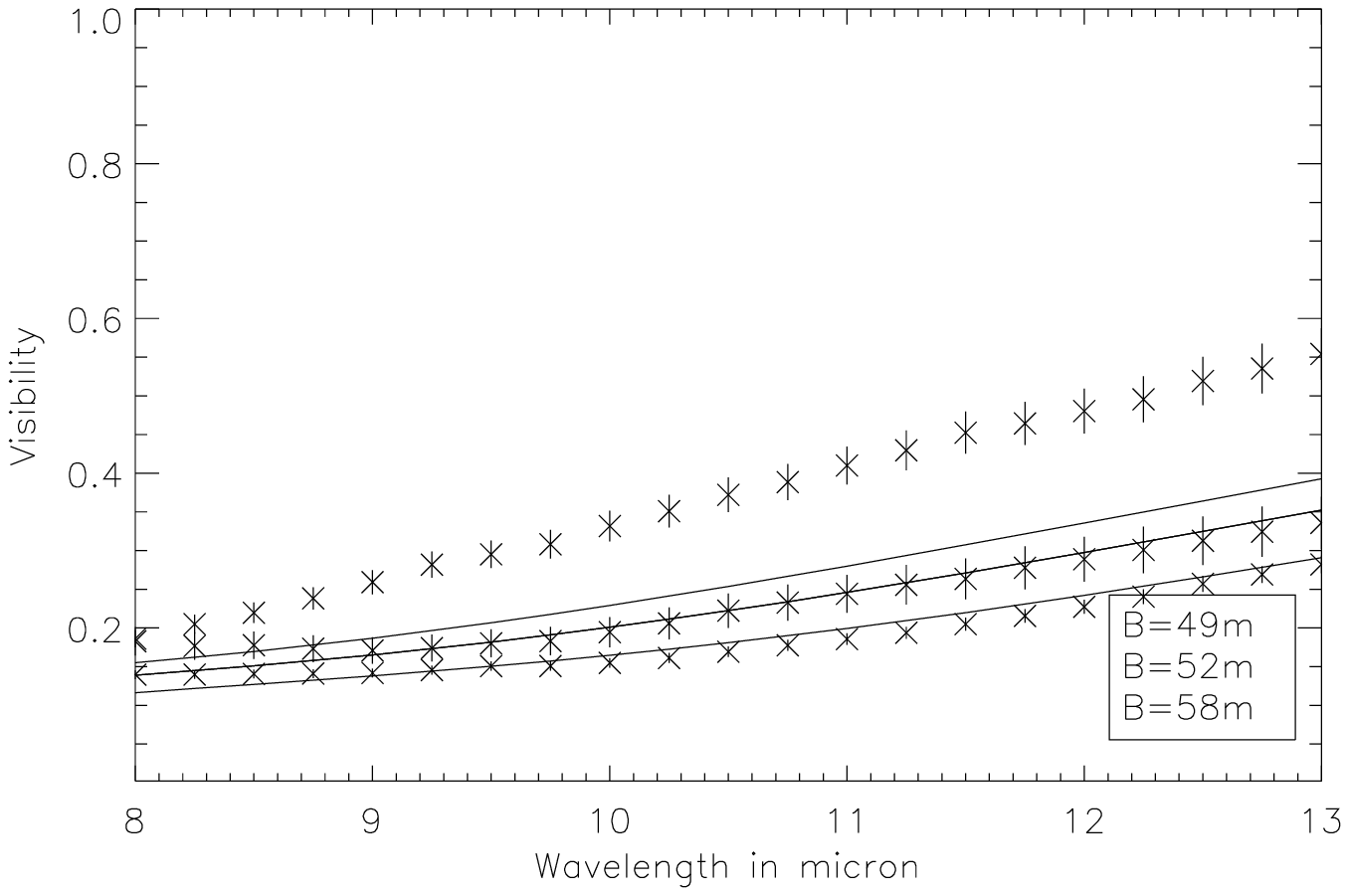}
\caption[]{Same as in Fig.\ref{fig:models} for days 110 and 145. At these dates the presence of a second shell is suspected, and indeed no synthetic visibility curves are found to agree within the full N band spectral range with the high SNR data. The last model is an alternative using a different power law dependence  for the density distribution i.e. $r^{-1.1}$ and not $r^{-2.}$ as for the other models. For further details see the text. 
\label{fig:models2}}
\end{figure*}

\subsection{The dust formation event}
Before the dust formation event, the K band continuum is expected to be dominated by  free-free 
emission from the nova wind \citep{1988ARAA..26..377G}. We have a single  K band visibility curve 
with AMBER at t=23d using two telescopes that tightly constrains the pseudo-photosphere size before 
or near the dust formation event. The K band visibility is 0.96$\pm$0.12 and the projected baseline is 71m. This implies a compact source that is unresolved given our large error bar at this date and a typical diameter smaller than approximately 1mas. This upper limit translates to a linear size of 1-2AU respectively for a corresponding distance value
of 1-2 kpc.

The AMBER spectrum does not show any rising continuum on this date in line with the Mt\,Abu spectroscopic observations recorded two days earlier, as can be seen in Fig.\ref{fig:IRspectra}, although 
a small contribution from an optically thin, already extended dust shell in formation cannot be excluded at this date. We note also that the limited spectral resolution does not allow us to isolate the emission lines that are still observable at that stage of the outburst (see Fig.\ref{fig:IRspectra}), albeit with a weak impact (contribution of a few percents).

Rapidly, the dust forms a dust shell optically thick from the visible to the N band, and a uniform disk is a good first approximation for the source (see \ref{sec:models}). The MIDI visibilities on day 36 are relatively inaccurate, followed on day 46 by better constrained observations, in close agreement with the AMBER observations. On day 90, the MIDI spectrally dispersed visibilities seem better fit by a Gaussian model than by a uniform disk, although it is difficult to disentangle between these models given the accuracy of the measurements.

On t=110 day, a difficulty arises in interpreting  the 35m baseline visibility curve which exhibits a slope that departs significantly from a uniform disk or Gaussian model. In an attempt to solve this issue, we interpret this effect by introducing a new unresolved component representing about 10-15\% of the flux. The residual of the curve is well fitted with a Gaussian with a FWHM of 27mas or a uniform disk of 43mas. This hypothesis is validated in view of the three visibility curves secured at t=145d (see Fig.6) with baselines ranging from 49m to 59m. At this date the shell should be as extended as $\sim$50mas, and with these baselines, the shell should be almost fully resolved. One can see that the three curves are not at the low level expected and converge to a level of $V\sim0.15$ at 8$\mu$m. From this level, and the slight differences between the visibilities from the different baselines (lower than error bars though), we can infer the presence of a new compact component of FWHM=13-17mas, contributing to $\sim$10-20\% of the N band flux. An alternative interpretation of the t=110d single visibility curve would be possible, but the presence of a new dust shell is established  beyond any doubt from the t=145d data.

\subsection{The shell expansion and distance determination}
From the angular diameters provided in Table\,2, one can infer the expansion rate of the main shell by interpolating the points with a linear expansion curve. This can be done at several chosen wavelengths
viz.  at 8, 10 and 13$\mu$m; each of them is subject to slightly different errors and bias. The first observations at t=36d are very noisy and the observations at t=110d and t=145d suffer from contamination intoduced by the presence of  the second shell that has to be carefully taken into account. This leads to a large uncertainty of the curves. We have tested three methods for determining the linear fit of the expansion rate based on three sets of data. The first one considers only the data points taken before t=110d in order to avoid the points affected by the appearance of the second shell. The second set includes all the visibility measurements, assuming simple uniform disk models, while the third data set uses the diameters of the main shell corrected for the presence of a compact source (indicated with an asterix in Table 2). 
The curves based on the second set show that the time of ejection is much before the  discovery of the outburst: this is unrealistic and shows that the data recorded at t=110d and t=145d must be corrected for the presence of a compact additional component. At 13$\mu$m, the curves based on the first data, and on the full corrected dataset agree fairly well, while at 8$\mu$m, there is a large discrepancy. This difficulty is related to the large error bars on the first measurement at t=36d.

The first conclusion is that the probable time of ejection is closer to the maximum of brightness than the date of discovery, with a computed mean date of t$_{ejec}=10.5\pm7$d for the zero crossing of the expansion curve. 
The mean expansion rate is determined with an accuracy better than 10\% at 0.35 $\pm$ 0.03 mas day$^{-1}$. As often reported, the main source of error for the distance determination comes from estimates of the dust shell velocity. The distance inferred from the expansion, assuming a velocity of the dust shell of 500$\pm$100km.s$^{-1}$ is 1.6$\pm$0.4kpc. This value is in rough agreement with the estimates based on the MM RD relations.


\subsection{Testing the spherical symmetry hypothesis}
\label{sect:sym}
The resolved remnants of novae ejecta  yield clues that nova shells often depart from spherical symmetry. This may point either to the fact that the eruption is intrinsically not spherically symmetric or that these large scale observations probe the interaction of the shell with its nearby interstellar environment. Recently, the spatially resolved observations the recurrent nova RS\,Oph have shown that the ejection was highly asymmetric, rapidly forming a nebula characterized by two prominent lobes and a dense equatorial waist \citep{2007ApJ...665L..63B, 2007A&A...464..119C, 2006Natur.442..279O}. This spatial complexity was visible from the very beginning of the outburst \citep{2007A&A...464..119C}. 

The sparse and time diluted VLTI observations provide little information on the shape of the source. No simultaneous use of three telescope was possible preventing closure phase spectra to be recorded. None of the 12 MIDI differential phases departs significantly from zero, with an error bar lower than 5$^\circ$. This eliminates some geometries such as a binary signal with moderate flux ratio but does not provide any constraint on the shape of a centrally-symmetric flux source. The last set of visibilities recorded at t=145d provide an indication of departure from sphericity that seems significant with regards to the error bars. Three baselines were used, covering about 75$^\circ$ with similar projected lengths. The baselines at PA=93$^\circ$ and 129$^\circ$ can be fitted with the same double shell model, while another set of parameters, involving a smaller shell is necessary to explain the level of the baseline at PA=166$^\circ$. However, a strong offset of this visibility curve is possible considering that this point was recorded at the end of the night with an airmass of 2 and in HIGH\_SENS mode (i.e with separated photometry), while the other two observations records were performed using the SCI\_PHOT mode (i.e. with simultaneous photometry).

In conclusion, the hypothesis that the ejection might have been aspherical is not constrained by the VLTI data and in absence of further information a spherical geometry will be considered. As a consequence, it is not possible to apply the corrections proposed in \citet{2000PASP..112..614W} for taking into account the asphericity of the source in the computation of the distance.

\section{Fitting the data with {\tt DUSTY} models}
\label{sec:models}
The use of uniform or Gaussian disks models has to be justified by a more physical model of the expanding dust shell. Uniform disks are probably the most appropriate to describe the optically thick dust shell, while the Gaussian distribution may fit better the latest stages when the shell dilutes and become optically thin. In this section we try to infer the parameters of the dust shell by using a spherical radiative transfer code. 

{\tt DUSTY} is a public domain simulation code solving the problem of radiation transport in a circumstellar dusty environment by analytically integrating the radiative transfer equation in plane-parallel or spherical geometries \citep{1999astro.ph.10475I,1997MNRAS.287..799I}. The code utilizes the self-similarity and scaling relation of the radiatively heated dust so that the shell is efficiently characterized by its optical depth. This means that 'absolute' values (e.g. luminosity and dust shell dimensions) are not uniquely determined by the transfer problem and must be inferred by external constraints - in our case, the recorded MIDI visibilities and spectra. 
The synthetic spectrally-dispersed visibility profiles throughout the N band
(7.5-–13$\mu$m) are generated using a set of 20 wavelengths and compared with the MIDI visibilities for
each baselines. At t=45-46d, the 2.2$\mu$m K band wavelength is also included to account for the AMBER measurement. 

Some assumptions are implied for the numerical simulations with DUSTY:\begin{itemize}
\item Spherical symmetry. This point is weakly constrained by our observations as discussed in Sect.\ref{sect:sym}. 
\item The dust shell is homogeneous (i.e. not clumpy).

\item The source is point-like as viewed from the shell. This assumption is probably hardly validated in reality, since the central source in the first moments of the outburst is a pseudo-photosphere surrounded by the screen of the ejected material, in which molecules and dust can form. This assumption is even more restrictive when considering the effect of a complex density structure (i.e. multiple shells) in the emerging SED and visibilities (see Discussion). 

\item the temperature and the spectrum of the source is constrained by model fitting. First simple blackbody fits were used in the models, but free-free curves provided much better quality fits (see following section).

\item the dust is made of amorphous carbon. There is no other real option in absence of dust features in the MIDI spectra. The near-IR spectra exhibited some carbon lines suggestive of a CO nova \citep{2007CBET..866....1D} that disappeared rapidly to be replaced by  a featureless continuum.

\item  The grain size distribution follows the classical MRN \citep{MRN77} relation: $  n(a) \propto a^{-q}      \hbox{for} \quad a_{\rm min} \le a \le a_{\rm max}$. Only $q$ was considered as free parameter during the fitting process, while the values $a_{\rm min}$=0.03$\mu$m and $a_{\rm max}$=3$\mu$m are kept fixed.

\item the thickness of the dust shell (R$_{out}$/R$_{in}$) is mainly constrained by the visibilities. Inside this shell the density follows a $r^{-2}$ distribution as in a steady-state wind with constant velocity. We performed many tests and the effects of changing the power of the density are difficult to evaluate: the effect is small for a thin shell and is relatively large for the latest stages.

\end{itemize}

 \begin{table*}
\begin{caption}
{Model parameters for V1280 Sco for the different epochs using a single dust shell. The parameters (and in particular the luminosity) are scaled for a distance of 1.6kpc. A chronology of dust formation based on an approximate date of dust  formation (t=23d) estimated from the light curve is also shown. 
For t=145d, two models are proposed which are indicated by designations 145 and 145$^b$. The second one is more extended and illustrates the degeneracy between models with different shell thicknesses 
at this stage which is related to the uncertainties in the velocity of the fastest ejecta (see text for discussion).
}\label{table-model} 

\end{caption}
\begin{tabular}{lccccccc}\hline
&\multicolumn{6}{c}{Day after outburst} \\
Days after discovery & 36   & 45/46  & 90 & 110 & 145 & 145$^b$ \\
Since dust detection &  13  & 22/23  & 67 & 87 & 122 & 122\\
\hline 
\\
T$_*$ (K) & 8500$\pm$1000  & 10000$\pm$1000 & 10000$\pm$1000 & 10000$\pm$1000 &10000$\pm$1000&10000$\pm$1000\\
T$_d$ (K) & 1450$\pm$150  & 1550$\pm$150 & 1300$\pm$100& 1450$\pm$100& 1550$\pm$100&1700$\pm$100\\
L (L$_\odot$) & 3500 & 4400 & 6600  & 16000 & 8400 & 8400\\
q &2.1$\pm$0.4 & 2.1$\pm$0.4& 2.9$\pm$0.3 & 2.9$\pm$0.3  & 3.0$\pm$0.2 & 3.0$\pm$0.2\\
Y=1+$\Delta R$/$R$& 1.3$\pm$0.3 & 1.5$\pm$0.3 & 2.8$\pm$0.4   &3.0$\pm$0.5 & 5.0$^{+5}_{-1}$.&17$^{+3}_{-6}$\\
r$_{in}$ (mas)&3.4&5&6.&6.4&3.8&2.2\\
r$_{in}$ (AU)&5.4&8&9.6&10.2&6.1 &3.5\\
r$_{out}$ (mas)&4.2&6.8&16.7&19.3&29.1&37\\
$\tau_V$ & 3.$\pm$0.3 & 4.5$\pm$0.3 & 6.5$\pm$0.3& 5.4$\pm$0.3&6.2$\pm$1.&5.5$\pm$0.4\\
$\tau_{10\mu m}$ & 2.3 & 4.1 & 2.8& 2.3&2.1 & 1.5\\
$\tau_{100\mu m}$ &4.43 10$^{-2}$&6.42 10$^{-2}$&4.05 10$^{-2}$&3.27 10$^{-2}$&2.47 10$^{-2}$&1.9 10$^{-2}$\\
M$_{dust} (M_\odot)$&4 10$^{-8}$&1.4 10$^{-7}$&2.2 10$^{-7}$&2.2 10$^{-7}$& 1.1 10$^{-7}$ &1 10$^{-7}$\\

\hline 
\end{tabular}
\end{table*}

\subsection{Simulating a single shell}
The parameters able to provide satisfactory fits to the data are shown in Table.2. The most complete set of observations was obtained at t=45-46d for which we have, in addition to the JHK photometry, both MIDI and AMBER observations yielding similar estimates of the shell size in K and N bands. This is a strong indication that the dust shell was  optically thick  and geometrically thin at that time.

The information extracted from this limited exercise can be summarized as follow:\begin{itemize}
\item The central source SED must depart strongly from a black-body. Often, unrealistically low temperatures were required to provide good fits using black-bodies, in line with the comments reported for the modeling of other novae using similar SEDs \citep{1998ASPC..137..489H}\footnote{Note that in this paper the authors tried to improve the fits by using two blackbody curves at 9000K and 3000K in an attempt to split the source between the outburst core and its compact vicinity. }). Assuming that the central source's radiation is effectively converted into free-free emission appears as a good alternative as the inner zones of the ejecta completely absorb the flux from the hot WD and subsequently emit free-free emission \citep{1988ARAA..26..377G}. Indeed, free-free spectra significantly improved the quality of the fits. The temperature associated with free-free spectra ranges between 7500 to 10000K. This corresponds to the temperatures associated with regions which are ionized by hot central sources, like HII regions, planetary nebulae, Be star disks in which the gas temperature is
usually close to 10,000K (see Banerjee et al. 2001). Similar quality fits are obtained when changing from a low temperature source SED to a higher one if one increases the optical depth so that the temperature of the source is not well constrained.
\item No grain size distributions with values of $q$ above 3 were able to provide good fits to the SEDs\footnote{We do not consider the last model proposed for t=145d as a viable solution}. This implies that the distribution of sizes of the grains formed is weighted towards large grains, and this from the very beginning of the observations. We find values of $q\sim$2.1 for the earliest dates that agrees well with the range found in Evans et al. (2005), but slightly larger values around $q\sim$3 seem necessary afterward. This suggests that processes decreasing the size of the dust grains were dominant just after their formation. This behavior is opposite to the the one reported in \citet[Fig.8]{1997MNRAS.292..192E}
\item The very first moments of the dust shell formation were caught by the VLTI interferometer when dust  continuously formed. As a consequence, the thickness of the shell (i.e. Y=1+$\Delta r /r$) was constantly increasing. However, we note that the SED fitting is not much affected by this parameter Y;  some family of solutions being found both for narrow (as low as $\sim 1.05$) as for wide values of Y. However, the observed visibilities constrain tightly the values reported in Table.\ref{table-model}. These large shell thicknesses are different than the narrow ones encountered for the modeling of a dust shell observed a long time after the outburst, whose value is set by the duration of the dust forming event and the internal velocity dispersion of the ejecta.
\item It was impossible to find solutions for all dates keeping the luminosity of the central source unchanged. This is related to several factors: the temperature and SED of the central source are not well constrained at these stages and the luminosity in the {\tt DUSTY} code determines the level of flux, but also the position of the inner radius of the shell. As such, it is an essential ingredient for the fitting of both spectra and visibilities. The difficulties encountered with this parameter illustrate the limits of the model (see Sect.\ref{sec:limits}),
\item As expected, single shell models have the highest difficulties in accounting for the last visibilities recorded at t=145d although  some models surprisingly are able to provide visibilities that mimic the plateau observed at the 0.15 level. One has to extend the shell thickness to unrealistically high values of Y which correspond to the outer radii reaching 30-40mas. This model probably bears some resemblance with  reality in the sense that the inner radius and its vicinity accounts for the new shell while the outer parts of the shell provides the screening of the first shell.
\end{itemize}
We also note that the spectral shape of the MIDI visibilities recorded at t=36d and t=46d are not smooth whereas the later visibilities exhibit a behavior more in line with the models. The obvious explanation is that these measurements are very noisy as the source was close to the sensitivity limit of the 1.8m Auxiliary Telescopes. This argument is surely valid for the high visibilities recorded from 10$\mu$m to 13$\mu$m. Nevertheless, the quality of the data is slightly better in the short wavelength range, and the visibilities recorded at t=36d are lower than expected by the model or by the trend of the expansion curves shown in Fig.\ref{fig:expansion}. A physical interpretation can be tentatively proposed, in the form of an additional opacity from molecular bands of various species, that contributes to screen the central source differently depending on the wavelength, although the molecules responsible from such a signal inside carbon rich ejecta are not well defined. It is not possible to state precisely whether this signal is effectively seen here, but this potentiality must be investigated in the future. We also note that the smoothness of the high SNR visibility curves obtained after t=90d provide an upper limit on the clumpiness of the shell although this limit excludes only the presence of a few, bright and mas-scale dust structures in the shell.

\subsection{Complex density profiles}
From day 110 on, the simple single shell models are no longer able to account for the shape of the visibility curves while being still able to fit the SEDs. Changing the power of the density profile for the full shell does not bring any significant improvement, although we suspect that the solution may reside in a complex density law that would reflect the variations of the rate of dust formation with time. As shown previously, there is some evidence in the latest visibilities recorded that some N band emission originates from more compact regions than the extended shell and we tried to model this by inserting a second shell in the DUSTY code. We first attempted to use a tabulated profile with two discrete shells but this kind of density profiles is not suited for the code. We also tried to assume a monotonic $r^{-2}$ density profile with density enhancements of the kind used in \citet{2001A&A...369..142B}. However, the monotonic density profile dominates the visibilities and the number of new parameters is large. We finally tried a third approach, still far from being self-consistent. The radiative transfer in the second (young) shell is first computed, and a SED and the visibilities are stored. The parameters of this shell are poorly constrained. The outcoming flux is then injected into the parameters in a second run in which the first extended shell is included although we stress that {\tt DUSTY} takes into account only the {\it shape} of the injected SED. There is no way to constrain the parameters of this shell but we note that the parameters of the t=36d shell might represent a good approximation given the estimations of the compact source size. The visibilities generated are biased since the extension of the inner shell is not taken into account by the code in the second run, and the flux contribution of the first shell has to be carefully taken into account for computing the flux ratio between the two shells. 
The last option, used for the t=145d models, was to test different density profiles from $r^{-3}$ to $r^{-1}$. A family of approximately good models can be found and we propose one of them in the last column of Table.\ref{table-model}. This example is an extreme case: the velocity implied by the extension  of the outer radius is about two times the rate inferred from the Uniform Disk models and the temperature is very high. Moreover, the value of $q$ is most probably overestimated as the SED fit is obviously less satisfactory. Nevertheless, it is worth noting that the mass inferred for this shell is in close agreement with the first option.
At some point in the evolution of the dust forming nova envelope, it is certain that the Center to Limb Variation (CLV) has evolved in a complex manner not accounted for by the simple models presented in this paper. Defining physically a 'second' shell forming event might be an over-interpretation of the data. Given the amount of open issues and the number of free parameters involved, the possibilities are almost infinite while the data remains fairly limited. We did not try to investigate this point further - it deserves an extensive study which is not in the scope of the paper. 

\subsection{Mass of the shell}
The {\tt DUSTY} models can provide a good estimate of the mass of the shell and hence of the dust formation rate during the event. The mass in the circumstellar dust shell was computed for each of the models following the relation for a r$^{-2}$ power-law dust density distribution \citep{2006ApJ...644.1171S}: 
$$M_{dust}=4 \pi R_{in}^2 Y (\tau_{100}/\kappa_{100})$$
 where $R_{in}$ is the inner radius of the dust shell (see Table \ref{table-model}), Y is the relative shell thickness (1+$\Delta R$/$R$), $\tau_{100}$ is the shell optical depth at 100$\mu$m and $\kappa_{100}$ is the dust mass absorption coefficient at 100$\mu$m. The value of $\kappa_{100}$ depends on the size of the grains, that varies in our models. We have used the values computed for amorphous carbon, that range between 58$cm^2 g^{-1}$ for $q$=2.1 (mean size 0.006$\mu$m) to 66$cm^2 g^{-1}$ for $q$=3 (mean size 0.003$\mu$m).
 
One can also get an estimation (with large errors) of the dust formed per day in between each observation. From the first shell modeled, one gets an amount of 3.1 10$^{-9} M_\odot$ day$^{-1}$, while between t=36d and t=45d, it reaches 7.4 10$^{-9} M_\odot$ day$^{-1}$. Between t=45d and t=110d, the dust formation rate is 1.8 10$^{-9} M_\odot$ day$^{-1}$, suggesting the dust forming process slowed down in that time interval, in line with the null forming rate estimated between t=90d and t=110d. Given the uncertainties of the data, the modeling and the date when dust began to form, these estimates have to be taken with some caution.
The lower amount of mass found at t=145d is also less significant than the other estimates due to the difficulty in finding visibilities that match the spectral shape of the observations. Performing a modeling by means of two distinct shells increases the amount of dust considerably.

It is probable that dust formation continued for several weeks before the rebrightening of the source in October-November 2007 \citep{2007CBET.1099....1M}. If one takes a mean value of dust forming rate of 5 10$^{-9} M_\odot$ day$^{-1}$, and a canonical gas-to-dust ratio of 150 (but this number can be much larger, see \citet{1988ARAA..26..377G}) then a dust shell of 1 10$^{-4} M_\odot$ is generated in about 130 days. V1280\,Sco could most probably have ejected a mass of material that  exceeds this number during the 200-250 days of the dust shell presence.

\subsection{Limitations of the model}
\label{sec:limits}
The assumptions for a good application of the DUSTY code are not necessarily fulfilled by V1280\,Sco. Aspects regarding the symmetry of the source have already been discussed in Sec.\ref{sect:sym}. We concentrate here on the hypothesis that the central source is considered as a point. The study of  nova V705\,Cas \citep{2005MNRAS.360.1483E}, using the same code, lies well within the limits imposed by this criterion.  Nova V705\,Cas was observed 250 days after the outburst when the central source was very hot and compact as seen from a detached, spatially and optically thin shell. This is not the case for V1280\,Sco  in the earliest stages when the dust shell was very close to the central source whose pseudo-photosphere is relatively cold and extended. The pseudo-photosphere is most probably surrounded by a diffuse free-free emission zone and farther away by a molecular atmosphere close to the dust forming region leading to a complex Center-to-Limb Variation profile. When the dust shell is optically thin, it is transparent to some emission coming from the internal regions that are not point-like and can be resolved by the interferometer, and which carry their own budget of correlated flux that is far from being taken into account in our models. 

In addition, there is a more fundamental limitation to the use of the {\tt DUSTY} code for the distance determination. We consider modeling a dust shell in rapid expansion. The central source is supposed to rapidly increase its temperature at constant bolometric luminosity. As discussed above, the radius of the shell is not defined in an absolute manner, but by setting a temperature for the central source. Intuitively, we could guess that as the shell expands its temperature will steadily decrease, but this is not observed as can be seen in Fig.\ref{fig:MIDI_flux}. This is the well-known 'isothermal' behavior described in Evans et al. (2005, see references there-in) and it directly biases the shell parameters presented in this work. The radius of the dust shell of V705 Cas is found to remain unchanged within the error bars between their epoch 1 (t=253d) and epoch 2 (t=320d) observations, while a 25\% increase is expected by the natural expansion of the shell. This is due to the fact that the shell radius in our {\tt DUSTY} models is set by the shell temperature and luminosity of the source and does not physically account for the dust facing the hot source (in particular any screening from the gas between the source and the dust). This internal inconsistency prevents us from hoping for a more reliable distance estimate using the model as long as the inner radius is not self-consistently determined in the model accounting for the source characteristics, the shell velocity and an accurate density law. We therefore consider the distance estimates based on the Uniform disk models as the most accurate one.

\section{Discussion}
\subsection{Dust shell expansion rate and distance estimation}
This is the first time that the dust shell of a classical nova is spatially resolved starting from the onset of the shell formation  up to the point it becomes close to  optically thin. This is a new and promising technique for providing dust-shell expansion rates, and thus distances, for interesting novae that are usually much fainter and almost featureless during the time that the dust shell is optically thick.
The present data set can provide a direct estimate of the distance to V1280\,Sco, a target for which a direct application of the MMRD relations is rendered difficult by the the dust formation event that seem to have occurred  {\it before} the light curve dropped below the defined thresholds $t_2$ and $t_3$.  

However, the applicability of such observations must be discussed. The first point, already discussed in \ref{sect:sym} is that the sphericity of the source is not established from our observations. The second point is that while  VLTI monitored the expansion of the dust shell, the velocities used to infer the distance were measured from visible line measurements. It is not sure that these velocity measurements trace the material that formed dust considering the observational evidence that the filling factor of the ejecta can be far from unity \citep{2006A&A...459..875E}. \citet{1988ARAA..26..377G} provide arguments in favor of the use of the determination of expansion parallaxes in the infrared. The dust forms preferentially in the densest regions of the ejecta, and may follow a different (slower) velocity law than the low-density wind itself. Yet, the consistency of the P Cygni absorption positions detected in many lines in optical and near-IR implies that the velocity of the dust forming shell is probably well represented by this value, and the error bar quoted in this paper is a conservative estimate. For the velocity of 575 km s$^{-1}$ reported in \citet{das2008}, the distance scales to $\sim$1.9kpc. This upper estimate must probably be favored in our interpretation.
Thirdly, uniform disks are a good approximation for describing the intensity distribution as long as the shell remains optically thick, but simple models are no longer suitable when the shell grows and the optical thickness in the N band gets low. The data recorded when the shell was optically thick are best estimators of the distance despite their lower accuracy.

A distance estimation based on the dust expansion parallax has some advantages compared to that using the expansion of the fireball in the near-IR as reported in \citet{2007ApJ...669.1150L,{2007ApJ...658..520L}}. The near-IR methods follows the fireball from its very first moments to the point where the the free-free emission is no longer optically thick. What remains at this stage is a compact, optically thick, shrinking source and an extended free-free emitting halo, rapidly over resolved by the interferometric technique. Moreover, the time to react to obtain observations  has to be fast, especially in the early stages of the outburst. In contrast in the mid-IR, the dust shell dominates the flux for a very long time, and the contribution of the hot core, even though not negligible, has less influence on the result. Nevertheless, the near-IR free-free expansion parallax technique has the advantage of being applicable to all novae and not only the ones that form dust. 

Our estimate for the time of ejection is subject to some uncertainty, but it seems probable that the material constituting the dust was launched after the beginning of the outburst, near the time of maximum light. It does not mean that material was not ejected earlier to this - onset of the  eruption is evidenced by P-Cygni profiles seen almost as early as t$_0$ and which lasted at least for 15 days thereafter: there was indeed a fireball expansion in the early stages. Why did dust not form in these early ejected layers? The temperature of the pseudo-photosphere of the firewall is a key parameter, since it was probably getting colder near and after the peak of visible light. Also, the mass-loss rate was most probably not constant and might have peaked close to the optical maximum: the dust formation rate might directly reflect this, although we can wonder whether the second event occurring at t=110d also reflected a sudden increase of the ejected mass.

\subsection{Multiple dust formation events}
The modeling using simple spherical shell models does not allow to clearly conclude on the formation of a second shell from t=110d on, but the light curve and the visibilities recorded at t=145d support this hypothesis.
\citet{2002AIPC..637..270R} proposed that the stability and hence the ultimate fate of the grains is primarily  determined by the degree to which they are annealed by the nova's ultraviolet radiation field. Shore \& Gehrz developed this idea arguing that agglomeration of atoms by dust nuclei proceeds kinetically through induced dipole reactions in a partially ionized medium. Ionization of a cluster increases once the ejecta become transparent in the ultraviolet which triggers runaway grain growth. We can wonder to which extent this mechanism intervenes in the formation of multiple 'shells' (i.e. large variations of dust formation rate with time). In the case of V1280\,Sco, signs of increase of the optical depth due to dust appears very early in the light curve, implying that, once nucleation had occurred, the physical conditions for grains growth (density and thus mass-loss rate and wind speed, temperature of the central source, and perhaps level of soft UV radiations) were optimum. The second shell must also have formed  close to the central star and independent  of the presence of the first shell. Therefore, its appearance can be considered as independent of the evolution of the first shell. Did the material in between the outbursting source and the dust forming region dilute below a level in which soft UV radiation emerged another time, triggering a new dust nucleation? A more likely hypothesis is an increase of the mass-loss rate (for unknown reasons) or perhaps a change to more favorable chemical conditions \citep{2004MNRAS.347.1294P} in conjunction with the hardening of the central source flux. If a 10-15mas new shell, due to a sudden increase of mass-loss rate, is observed at t=145d then this material was generated about 35-40 days before the observations (still assuming a velocity of 500km.s$^{-1}$) i.e. around t=110d  when the second light maximum occurred. We can then make the hypothesis that this second maximum is not only a consequence of the decrease of the expanding dust optical depth but is also related to an increase of the luminosity and mass-loss rate of the central source. One can note in Table.\ref{table-model} that the luminosity required to account for the observed data at t=110d is notably larger than at other epochs.


\section{Conclusions}
High spatial resolution observations of the very first moments of a nova outburst promise to provide invaluable information on the physical processes operating in the vicinity of the central remnant. The
VLTI observations reported here, the first  of their kind,  show that spectro-interferometric observations can localize accurately the dust forming regions allowing a 'virtual' in-situ study of the dust forming event. 

These observations present interesting information on the dust shell formed around the classical nova V1280 Sco. But they still lack the minimum $uv$ coverage needed to check the sphericity of the source. There was scope for the near-IR photometry and spectroscopic data obtained by \citet{das2008} from this bright source to be complemented by intensive VLTI/AMBER observations. The source was bright enough during the 30 first days to be studied by VLTI with a spectral resolution of 1500 in H and K, providing potentially direct constraints on the spatial extent of the line forming regions and their kinematics (such as in \citet{2007A&A...464..119C}). Our relatively recent involvement in this field and inadequate manpower prevented our team from undertaking  these potentially exciting observations - but scope exists for monitoring one or two bright novae per year by this facility. This field, and studies of a similar kind, should  greatly benefit from the outcome of  2nd generation instrumentation efforts of the VLTI. The mid-IR instrument MATISSE \citep{2006SPIE.6268E..31L} is foreseen to recombine the light from 4 telescopes, providing in one shot six visibility measurements and three closure phases that should provide at one go 
a detailed view of a dust forming shell.

\begin{acknowledgements}
The VLTI staff is warmly thanked for their intensive efforts in  efficiently operating the facility. By constantly communicating the latest observational results to our team, they helped us to plan and understand the observations. This research has made use of the AFOEV database, operated at CDS, France. We also acknowledge with thanks the variable star observations from the AAVSO International Database contributed by observers worldwide.  The blog of the KANATA 
1.5-m telescope at Higashi-Hiroshima Observatory, Japan has also provided valuable information for the planning and preliminary understanding of our observations. The research work at Physical Research Laboratory is funded by the Department of Space, Government of India.
\end{acknowledgements}


\begin{thebibliography}{}
\bibitem[Banerjee et al.(2001)]{2001A&A...380L..13B} Banerjee, D.~P.~K., 
Janardhan, P., \& Ashok, N.~M.\ 2001, \aap, 380, L13 
\bibitem[Barry et al.(2008)]{barry2008} Barry, R.K. Danchi, W.C., Traub, W.A. 
et al.\ 2008, ArXiv e-prints, 801, arXiv: 0801:4165
\bibitem[Bl{\"o}cker et al.(2001)]{2001A&A...369..142B} Bl{\"o}cker, T., 
Balega, Y., Hofmann, K.-H., \& Weigelt, G.\ 2001, \aap, 369, 142 
\bibitem[Bode et al.(2007)]{2007ApJ...665L..63B} Bode, M.F., Harman, 
D.J., O'Brien, T.J. et al.\ 2007, \apjl, 665, L63 
\bibitem[Chesneau et al.(2007)]{2007A&A...464..119C} Chesneau, O., Nardetto, N., Millour, F. et al.\ 2007, \aap, 464, 119 
\bibitem[Cassatella et al.(2004)]{2004A&A...420..571C} Cassatella, A., 
Lamers, H.J.G.L.M., Rossi, C. et al.\ 2004, \aap, 420, 571 
\bibitem[Das et al.(2008)]{das2008} Das, R.K., Banerjee, D.P.K., Ashok, N.M. and Chesneau, O., 2008, MNRAS, accepted
\bibitem[Das et al.(2007)]{2007CBET..866....1D} Das, R.K., Ashok, N.M., 
\& Banerjee, D.P.K.\ 2007, Central Bureau Electronic Telegrams, 866, 1 
\bibitem[Das et al.(2007)]{2007CBET..864....1D} Das, R.K., Ashok, N.M., 
\& Banerjee, D.P.K.\ 2007, Central Bureau Electronic Telegrams, 864, 1 
\bibitem[della Valle \& Livio(1995)]{1995ApJ...452..704D} della Valle, M., 
\& Livio, M.\ 1995, \apj, 452, 704 
\bibitem[Deroo et 
al.(2007)]{2007A&A...467.1093D} Deroo, P., van Winckel, H., Verhoelst, T. et al.\ 2007, \aap, 467, 1093 
\bibitem[Ederoclite et al.(2006)]{2006A&A...459..875E} Ederoclite, A., et 
al.\ 2006, \aap, 459, 875 
\bibitem[Evans et al.(1996)]{1996MNRAS.282.1049E} Evans, A., Geballe, 
T.R., Rawlings, J.~M.~C., \& Scott, A.~D.\ 1996, \mnras, 282, 1049 
\bibitem[Evans et al.(1997)]{1997MNRAS.292..192E} Evans, A., Geballe, 
T.R., Rawlings, J.M.C. et al.\ 1997, \mnras, 
292, 192 
\bibitem[Evans et al.(2005)]{2005MNRAS.360.1483E} Evans, A., Tyne, V.H., 
Smith, O. et al.\ 2005, 
\mnras, 360, 1483 
\bibitem[Gehrz(1988)]{1988ARAA..26..377G} Gehrz, R.~D.\ 1988, ARA\&A, 26, 
377 
\bibitem[Gehrz et al.(1995)]{1995ApJ...448L.119G} Gehrz, R.D., Greenhouse, 
M.~A., Hayward, T.~L. et al.\ 
1995, \apjl, 448, L119 
\bibitem[Gehrz et al.(1998)]{1998PASP..110....3G} Gehrz, R.~D., Truran, 
J.~W., Williams, R.~E., \& Starrfield, S.\ 1998, \pasp, 110, 3 
\bibitem[Harrison et al.(1998)]{1998ASPC..137..489H} Harrison, T.E., 
Johnson, J.J., Mason, P.A., \& Stringfellow, G.~S.\ 1998, ASP Conference Series, 137, 489 
\bibitem[Ivezic et al.(1999)]{1999astro.ph.10475I} Ivezic, Z., Nenkova, M., 
\& Elitzur, M.\ 1999, ArXiv Astrophysics e-prints, arXiv:astro-ph/9910475 
\bibitem[Ivezic \& Elitzur(1997)]{1997MNRAS.287..799I} Ivezic, Z., \& 
Elitzur, M.\ 1997, \mnras, 287, 799 
\bibitem[Lane et al.(2007b)]{2007ApJ...669.1150L} Lane, B.F., Retter, A., 
Eisner, J.A., et al.\ 2007, \apj, 669, 1150 
\bibitem[Lane et al.(2007a)]{2007ApJ...658..520L} Lane, B.~F., Sokoloski, J.L., Barry, R.K. et al.\ 2007, \apj, 658, 520
\bibitem[Lopez et al.(2006)]{2006SPIE.6268E..31L} Lopez, B., Wolf, S., Lagarde, S. et al.\ 2006, 
\procspie, 6268, 
\bibitem[Marshall et al.(2006)]{2006A&A...453..635M} Marshall, D.~J., 
Robin, A.~C., Reyl{\'e}, C., Schultheis, M., \& Picaud, S.\ 2006, \aap, 
453, 635 
\bibitem[Mathis, Rumpl \& Nordsieck(1977)]{MRN77}  Mathis J.S., Rumpl W. \& Nordsieck K.H. 1977, ApJ, 217, 425
\bibitem[Millour et al.(2007)]{2007arXiv0705.1636M} Millour, F., et al.\ 
2007, ArXiv e-prints, 705, arXiv:0705.1636 
\bibitem[Monnier et al.(2006)]{2006ApJ...647L.127M} Monnier, J.~D., et al.\ 
2006, \apjl, 647, L127 
\bibitem[Munari et al.(2007b)]{2007CBET.1099....1M} Munari, U., Siviero, A., 
Henden, A. et al.\ 2007, Central Bureau Electronic Telegrams, 1099, 1 
\bibitem[Munari et al.(2007a)]{2007CBET..852....1M} Munari, U., Valisa, P., 
Dalla Via, G., \& Dallaporta, S.\ 2007, Central Bureau Electronic 
Telegrams, 852, 1 
\bibitem[Naito \& Narusawa(2007)]{2007IAUC.8803....2N} Naito, H., \& 
Narusawa, S.\ 2007, \iaucirc, 8803, 2 
\bibitem[O'Brien et al.(2006)]{2006Natur.442..279O} O'Brien, T.J., Bode, M.F., Porcas, R.W. et al.\ 2006, \nat, 442, 279 
\bibitem[Petrov et al.(2007)]{2007A&A...464....1P} Petrov, R.~G., Malbet, F., Weigelt, G. et al.\ 
2007, \aap, 464, 1 
\bibitem[Pontefract \& Rawlings(2004)]{2004MNRAS.347.1294P} Pontefract, M., 
\& Rawlings, J.M.C.\ 2004, \mnras, 347, 1294 
\bibitem[Quirrenbach et al.(1993)]{1993AJ....106.1118Q} Quirrenbach, A., Elias, N.M., Mozurkewich, D., et al.\ 1993, AJ, 106, 1118
\bibitem[Rawlings \& Evans(2002)]{2002AIPC..637..270R} Rawlings, J.~M.~C., 
\& Evans, A.\ 2002, Classical Nova Explosions, 637, 270 
\bibitem[Rudy et al.(2007)]{2007IAUC.8809....1R} Rudy, R.J., Lynch, D.~K., 
Russell, R.W., Woodward, C.E., Liebert, J., \& Cushing, M.\ 2007, 
\iaucirc, 8809, 1 
\bibitem[Rudy et al.(2003)]{2003ApJ...596.1229R} Rudy, R.J., Dimpfl, 
W.L., Lynch, D.K. et al.\ 2003, \apj, 596, 1229 
\bibitem[Sarkar \& Sahai(2006)]{2006ApJ...644.1171S} Sarkar, G., \& Sahai, 
R.\ 2006, \apj, 644, 1171 
\bibitem[Sch{\"o}ller et al.(2006)]{2006SPIE.6268E..19S} Sch{\"o}ller, M., 
et al.\ 2006, \procspie, 6268, 19 
\bibitem[Shore \& Gehrz(2004)]{2004A&A...417..695S} Shore, S.N., \& Gehrz, 
R.~D.\ 2004, \aap, 417, 695 
\bibitem[Tatulli et al.(2007)]{2007A&A...464...29T} Tatulli, E., Millour, F., Chelli, A. et al.\ 
2007, \aap, 464, 29 
\bibitem[Wade et al.(2000)]{2000PASP..112..614W} Wade, R.~A., Harlow, 
J.~J.~B., \& Ciardullo, R.\ 2000, \pasp, 112, 614 
\bibitem[Yamaoka et al.(2007)]{2007IAUC.8807....2Y} Yamaoka, H., Fujii, M., 
\& Naito, H.\ 2007, \iaucirc, 8807, 2 
\bibitem[Yamaoka et al.(2007)]{2007IAUC.8803....1Y} Yamaoka, H., Nakamura, 
Y., Nakano, S., Sakurai, Y., \& Kadota, K.\ 2007, \iaucirc, 8803, 1 

\end{thebibliography}
\end{document}